\documentclass{article}

\usepackage{microtype}
\usepackage{graphicx}
\usepackage{booktabs} 

\usepackage{hyperref}

\usepackage[accepted]{icml2023}

\usepackage{amsmath}
\usepackage{amssymb}
\usepackage{mathtools}
\usepackage{amsthm}

\usepackage{multirow}
\usepackage{subcaption}
\usepackage{tablefootnote}
\usepackage{enumitem}

\usepackage[capitalize,noabbrev]{cleveref}

\theoremstyle{plain}
\newtheorem{theorem}{Theorem}[section]
\newtheorem{proposition}[theorem]{Proposition}
\newtheorem{lemma}[theorem]{Lemma}

\theoremstyle{definition}
\newtheorem{definition}[theorem]{Definition}

\theoremstyle{remark}

\usepackage[textsize=tiny]{todonotes}

\newcommand{\proposed}{\textsf{CGIB}}
\newcommand{\proposedcont}{$\textsf{CGIB}_{\textrm{cont}}$}

\DeclareMathOperator*{\argmin}{arg\,min}

\usepackage{xcolor}
\usepackage{pifont}
\usepackage{graphicx}
\newcommand{\cmark}{\textcolor{blue}{\ding{51}}}%
\newcommand{\xmark}{\textcolor{red}{\ding{55}}}%

\icmltitlerunning{Conditional Graph Information Bottleneck for Molecular Relational Learning}

\begin{document}

\twocolumn[
\icmltitle{Conditional Graph Information Bottleneck for Molecular Relational Learning}




\begin{icmlauthorlist}
\icmlauthor{Namkyeong Lee}{kaist}
\icmlauthor{Dongmin Hyun}{postech}
\icmlauthor{Gyoung S. Na}{krict}
\icmlauthor{Sungwon Kim}{kaist}
\icmlauthor{Junseok Lee}{kaist}
\icmlauthor{Chanyoung Park}{kaist}
\end{icmlauthorlist}

\icmlaffiliation{kaist}{KAIST}
\icmlaffiliation{postech}{POSTECH}
\icmlaffiliation{krict}{KRICT}
\icmlcorrespondingauthor{Chanyoung Park}{cy.park@kaist.ac.kr}

\icmlkeywords{Machine Learning, ICML}

\vskip 0.3in
]



\printAffiliationsAndNotice{}  

\begin{abstract}
Molecular relational learning, whose goal is to learn the interaction behavior between molecular \textit{pairs}, got a surge of interest in molecular sciences due to its wide range of applications.
Recently, graph neural networks have recently shown great success in molecular relational learning by modeling a molecule as a graph structure, and considering atom-level interactions between two molecules.
Despite their success, existing molecular relational learning methods tend to overlook the nature of chemistry, i.e., a chemical compound is composed of multiple substructures such as functional groups that cause distinctive chemical reactions.
In this work, we propose a novel relational learning framework, called \proposed, that predicts the interaction behavior between a pair of graphs by detecting core subgraphs therein.
The main idea is, given a pair of graphs, to find a subgraph from a graph that contains the minimal sufficient information regarding the task at hand conditioned on the paired graph based on the principle of conditional graph information bottleneck.
We argue that our proposed method mimics the nature of chemical reactions, i.e., the core substructure of a molecule varies depending on which other molecule it interacts with. 
Extensive experiments on various tasks with real-world datasets demonstrate the superiority of \proposed~over state-of-the-art baselines.
Our code is available at \url{https://github.com/Namkyeong/CGIB}.
\end{abstract}

\section{Introduction}

\begin{figure}[t]
    \centering
    \includegraphics[width=0.75\columnwidth]{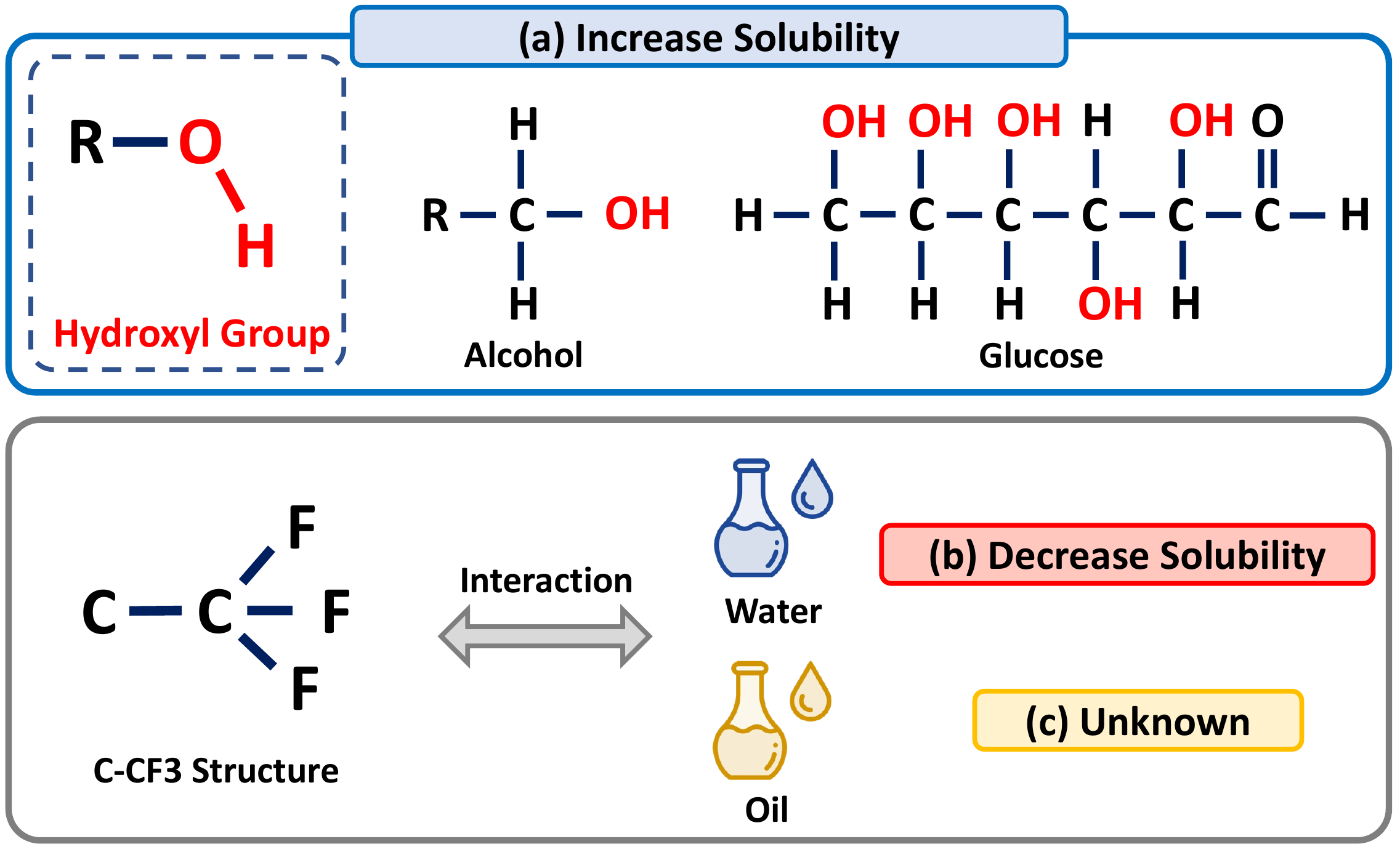} 
    \caption{(a) Molecules with hydroxyl group tend to have high aqueous solubility. (b) C-CF3 structure decreases the solubility of molecules in water. (c) However, it may not be crucial in determining the solubility of molecules in oil.}
    \label{fig1}
    \vspace{-4ex}
\end{figure}

Relational learning \cite{rozemberczki2021unified}, which aims to predict the interaction behavior between entity \textit{pairs}, got a surge of interest among researchers due to its wide range of applications, especially in molecular science, which is the main focus of this paper.
For example, predicting optical and photophysical properties of chromophores with various solvents (i.e., chromophore-solvent pair) is important in designing and synthesizing new colorful materials \cite{joung2021deep}.
Moreover, it is crucial to determine how medications will dissolve in various solvents (i.e., medication-solvent pair) and how different drug combinations will interact with each other (i.e., drug-drug pair) \cite{CIGIN, MIRACLE}.
Due to the expensive time/financial costs of exhaustively conducting experiments to test the interaction behavior between all possible molecular pairs \cite{preuer2018deepsynergy}, machine learning approaches have been rapidly adopted for relational learning in molecular sciences~\cite{joung2021deep,CIGIN,MIRACLE}.

In this paper, we propose a novel molecular relational learning framework inspired by the nature of chemistry: \textit{a chemical compound is composed of multiple substructures such as functional groups that cause distinctive chemical reactions}.
That is, a certain functional group is known to induce the same or similar chemical reactions regardless of other components that exist in the chemical, and thus considering functional groups facilitates a systematic prediction of chemical reactions and the behavior of chemical compounds \cite{book2014compendium,jerry1992advanced}.
For example, as shown in Figure \ref{fig1}(a), alcohol and glucose commonly contain hydroxyl group, which increases the polarity of molecules.
Thus, alcohol and glucose tend to have high aqueous solubility due to the hydroxyl group \cite{delaney2004esol}.
In this regard, it is crucial for a model to detect the core substructures of chemicals to improve its generalization ability.
However, detecting the core subgraph (i.e., substructure) of an input graph (i.e., chemical compound) is not trivial due to its complex nature \cite{alsentzer2020subgraph,meng2018subgraph}.

Recently, information bottleneck (IB) theory \cite{tishby2000information} has been applied to learning significant subgraphs of the input graph for explainable GNNs \cite{VGIB,PGIB,GSAT}, which provides a principled approach to determine which aspects of data should be preserved and which should be discarded \cite{pan2021disentangled}.
Specifically, GSAT \cite{GSAT} formulates the subgraph attention as an information bottleneck by learning stochastic attention that randomly drops edges and obtains a perturbed graph which is considered as an explanatory subgraph.
Moreover, VGIB \cite{VGIB} obtains a perturbed graph by selectively injecting noises into unnecessary node representations, thereby modulating the information flow from the original graph into the perturbed graph.

However, directly applying graph information bottleneck (GIB) into a relational learning framework is challenging since the complexity arises not only within a single graph, but also between graphs.
Specifically, there exist several meaningful subgraphs within a graph, and the importance of each subgraph varies depending on which other graph it interacts with. 
For example, while C-CF3 substructure plays an important role in decreasing the solubility of molecules in water \cite{purser2008fluorine} (Figure~\ref{fig1}(b)), it may not be crucial in determining the solubility of molecules in oil (Figure \ref{fig1}(c)). This implies that the importance of a substructure depends on the context.

To this end, we propose \textsf{C}onditional \textsf{G}raph \textsf{I}nformation \textsf{B}ottleneck (\proposed), a simple yet effective relational learning framework that predicts the interaction behavior between a \textit{pair} of graphs by detecting important subgraphs therein.
Our main goal is, given a pair of graphs $\mathcal{G}^1$ and $\mathcal{G}^2$, to detect the subgraph of $\mathcal{G}^1$ (i.e., $\mathcal{G}_{\mathrm{CIB}}^1$) that is crucial in determining the interaction behavior between $\mathcal{G}^1$ and $\mathcal{G}^2$.
Existing GIB-based approaches are designed for tasks that require a single input graph, and thus the core subgraph is learned solely based on the input graph itself \cite{PGIB,VGIB,GSAT}. On the other hand, as~\proposed~is designed for relational learning,
the core subgraph learned by \proposed~varies according to the pair of graphs at hand (i.e., the core subgraph of $\mathcal{G}^1$ varies depending on the paired graph $\mathcal{G}^2$).
Specifically, given a graph $\mathcal{G}^1$, CGIB learns its core subgraph $\mathcal{G}^1_{\mathrm{CIB}}$ that maximizes the mutual information between a pair of graphs (i.e., ($\mathcal{G}_{\mathrm{CIB}}^1$, $\mathcal{G}^2$)) and the target response (i.e., $\mathbf{Y}$) we aim to predict, 
while minimizing the mutual information between the graph $\mathcal{G}^1$ and its subgraph $\mathcal{G}^1_{\mathrm{CIB}}$ conditioned on its paired graph $\mathcal{G}^2$.
Moreover, based on the chain rule of mutual information, the conditional mutual information is minimized by injecting Gaussian noise into the node representations of $\mathcal{G}^1$, which controls the information flow between the graph and its subgraph $\mathcal{G}^1_{\mathrm{CIB}}$, while maximizing the mutual information between the subgraph $\mathcal{G}^1_{\mathrm{CIB}}$ and the paired graph $\mathcal{G}^2$.
By doing so, CGIB learns the subgraph $\mathcal{G}^1_{\mathrm{CIB}}$ that contains minimal sufficient information regarding both the paired graph $\mathcal{G}^2$ and the target value $\mathbf{Y}$.

Our extensive experiments on eleven real-world datasets on various tasks, i.e., molecular interaction prediction, drug-drug interaction prediction, and graph similarity learning, demonstrate the effectiveness and generality of \proposed~in relational learning problems.
{
Moreover, ablation studies verify that \proposed~successfully adopts the IB principle to relational learning, which is non-trivial.
A further appeal of \proposed~is its explainability, i.e., discovering the core substructure of the chemical compounds during chemical reactions, as shown in our qualitative analysis.
To the best of our knowledge,~\proposed~is the first work that adopts the IB principle to relational learning tasks.
}

\section{Related Work}
\subsection{Molecular Relational Learning}

Molecular relational learning can be categorized into two categories according to the target types, i.e., molecular interaction prediction and drug-drug interaction prediction.

\noindent \textbf{Molecular Interaction Prediction.}
In the molecular interaction prediction task, a model predicts the properties of chemicals induced by chemical reactions or properties of the reaction itself.
Delfos \cite{lim2019delfos} predicts the solvation free energy, which is directly related to the solubility of chemical entities, by using recurrent neural networks and attention mechanisms with SMILES sequence as the input.
CIGIN \cite{CIGIN} leverages message passing neural networks \cite{MPNN} and co-attention mechanism to encode the representation of atoms to predict the solvation free energy.
Moreover, CIGIN further enhances the interpretability of chemical reactions with co-attention map, which indicates the importance of interaction between atoms.
\citet{joung2021deep} predict diverse optical and photophysical properties of chromophores, which play a critical role in synthesizing new colorful materials, with the representations of chromophores and solvents that are from graph convolutional networks \cite{GCN}.

\noindent \textbf{Drug-Drug Interaction.}
In the drug-drug interaction task, the model classifies which type of interaction will occur between drugs.
Specifically, under the assumption that drugs sharing similar structures tend to share similar interactions,
\citet{vilar2012drug} and \citet{kastrin2018predicting} predict DDI by comparing the Tanimoto coefficient of drug fingerprints and exhibiting explicit similarity-based features, respectively.
MHCADDI \cite{deac2019drug} proposes a co-attentitive message passing network \cite{GAT} for polypharmacy side effect prediction, which, given a pair of molecules, aggregates the messages not only from the atoms inside a single molecule, but also all atoms in the paired molecule.
MIRACLE \cite{MIRACLE} casts the DDI task as a link prediction task by constructing a multi-view graph, where each node in the interaction graph itself is a drug molecular graph instance.

However, existing studies in molecular relational learning do not consider core substructures such as functional groups, which determine distinctive chemical characteristics.
Moreover, they are designed for a specific task, raising doubts on the generality of methods.
In this work, we propose a general framework that predicts the behavior of graph pairs by detecting core subgraphs therein.

\subsection{Graph Information Bottleneck}
Recent studies have introduced the IB principle to graph-structure data, whose irregular data structure incurs unique challenges in computing the mutual information 
\cite{PGIB,VGIB,GSAT,sun2022graph,GIB}.
Specifically, GIB \cite{GIB} extends the general IB principle for node representation learning by regularizing both structure and feature information.
GIB \cite{PGIB} studies the subgraph recognition problem by formulating a subgraph as a bottleneck random variable. It employs the Shannon mutual information to measure how compressed and informative the subgraph distribution is.
VGIB \cite{VGIB} further stabilizes the subgraph recognition process by injecting Gaussian noise into node representations, where the noise modulates the information flow from the original graph into the perturbed graph.
GSAT \cite{GSAT} obtains a subgraph by applying stochastic attention, which randomly drops edges with parameterized Bernoulli distribution.
Although the IB principle has been successfully applied to graph-structured data, previous studies have focused on tasks that require a single input graph, and thus have only taken into account a single graph for subgraph recognition, which limits their applicability to relational learning tasks.
To the best of our knowledge, \proposed~is the first work that adopts the IB principle for relational learning.

\section{Preliminaries}
In this section, we first formally describe the problem formulation including notations and the task description (Section \ref{Sec: Problem Formulation}). Then, we introduce definitions of IB and IB-Graph, which is an application of IB to recognize the core subgraph of an input graph (Section \ref{Sec: Information Bottleneck}).

\subsection{Problem Formulation}
\label{Sec: Problem Formulation}
\noindent \textbf{Notations.}
Let $\mathcal{G} = (\mathcal{V}, \mathcal{E})$ denotes a graph, where $\mathcal{V} = \{v_{1},\ldots, v_{N}\}$ represents the set of nodes, and $\mathcal{E} \subseteq \mathcal{V} \times \mathcal{V}$ represents the set of edges. $\mathcal{G}$ is associated with a feature matrix $\mathbf{X} \in \mathbb{R}^{N \times F}$, and an adjacency matrix $\mathbf{A} \in \mathbb{R}^{N \times N}$ where $\mathbf{A}_{ij} = 1$ if and only if $(v_i, v_j) \in \mathcal{E}$ and $\mathbf{A}_{ij} = 0$ otherwise.
We denote $I(X; Y)$ as the mutual information between random variables $X$ and $Y$:
\begin{equation}
	I(X; Y) = \int_{X} \int_{Y} p(x, y) \log \frac{p(x, y)}{p(x)p(y)} dx dy
\end{equation}

\noindent \textbf{Task: Relational Learning on Graphs.}
Given a set of graph pairs $\mathcal{D} = \{(\mathcal{G}_1^{1}, \mathcal{G}_1^{2}), (\mathcal{G}_2^{1}, \mathcal{G}_2^{2}), \ldots, (\mathcal{G}_n^{1}, \mathcal{G}_n^{2}) \}$ and the associated target values $\mathbb{Y} = \{\mathbf{Y}_1, \mathbf{Y}_2, \ldots, \mathbf{Y}_n \}$, our goal is to train a model $\mathcal{M}$ that predicts the target values of given arbitrary graph pairs in an end-to-end manner, i.e., $\mathbf{Y}_{i} = \mathcal{M}(\mathcal{G}_{i}^{1}, \mathcal{G}_{i}^{2})$.
The target $\mathbf{Y}$ is a scalar value, i.e., $\mathbf{Y} \in (-\infty, \infty)$, for regression tasks, while it is a binary class label, i.e., $\mathbf{Y} \in \{0, 1\}$, for (binary) classification tasks.

\subsection{Information Bottleneck}
\label{Sec: Information Bottleneck}
In machine learning, it is important to determine which aspects of the input data should be preserved and which should be discarded.
Information bottleneck (IB) \cite{tishby2000information} provides a principled approach to this problem, compressing the source random variable to keep the information relevant for predicting the target random variable while discarding target-irrelevant information.

\smallskip
\begin{definition} 
\textit{(Information Bottleneck)}
Given random variables $X$ and $Y$, the Information Bottleneck principle aims to compress $X$ to a bottleneck random variable $T$, while keeping the information relevant for predicting $Y$:
\begin{equation}
	\min_{T} -I(Y; T) + \beta I(X; T)
\label{eq: IB}
\end{equation}
where $\beta$ is a Lagrangian multiplier for balancing the two mutual information terms.
\end{definition}

Recently, IB principle has been applied to learning a bottleneck graph $\mathcal{G}_{\mathrm{IB}} = (\mathbf{X}_{\mathrm{IB}}, \mathbf{A}_{\mathrm{IB}})$ named IB-Graph for $\mathcal{G}$, which keeps minimal sufficient information in terms of $\mathcal{G}$'s properties \cite{PGIB,VGIB,GSAT}.

\smallskip
\begin{definition} 
\textit{(IB-Graph)}
For a graph $\mathcal{G} = (\mathbf{X}, \mathbf{A})$ and its label information $\mathbf{Y}$, the optimal graph $\mathcal{G}_{\mathrm{IB}} = (\mathbf{X}_{\mathrm{IB}}, \mathbf{A}_{\mathrm{IB}})$ discovered under the IB principle is denoted as IB-Graph:
\begin{equation}
	\mathcal{G}_{\mathrm{IB}} = \argmin_{\mathcal{G}_{\mathrm{IB}}} -I(\mathbf{Y} ;\mathcal{G}_{\mathrm{IB}}) + \beta I(\mathcal{G}; \mathcal{G}_{\mathrm{IB}})
 \label{eq:IBgraph}
\end{equation}
where $\mathbf{X}_{\mathrm{IB}}$ and $\mathbf{A}_{\mathrm{IB}}$ denote the task-relevant feature set and the adjacency matrix of $\mathcal{G}$, respectively.
\end{definition} 

Intuitively, graph information bottleneck (GIB) aims to learn the core subgraph of the input graph (i.e., $\mathcal{G}_{\mathrm{IB}}$), which discards information from the input graph by minimizing the term $I(\mathcal{G}; \mathcal{G}_{\mathrm{IB}})$, while preserving target-relevant information by maximizing the term $I(\mathbf{Y} ;\mathcal{G}_{\mathrm{IB}})$.

\section{Methodology}

In this section, we introduce our proposed method called \proposed, a novel relational learning framework that detects the core subgraph of an input graph based on the conditional mutual information.
First, we formally define the conditional graph information bottleneck and CIB-Graph (Section \ref{Sec: Methodology Conditional Graph Information Bottleneck}).
Then, we introduce the overall model architecture (Section \ref{Sec: Model Architecture}) followed by the overall optimization process of~\proposed~(Section \ref{Sec: Model Optimization}).

\begin{figure}[t]
    \centering
    \includegraphics[width=0.9\linewidth]{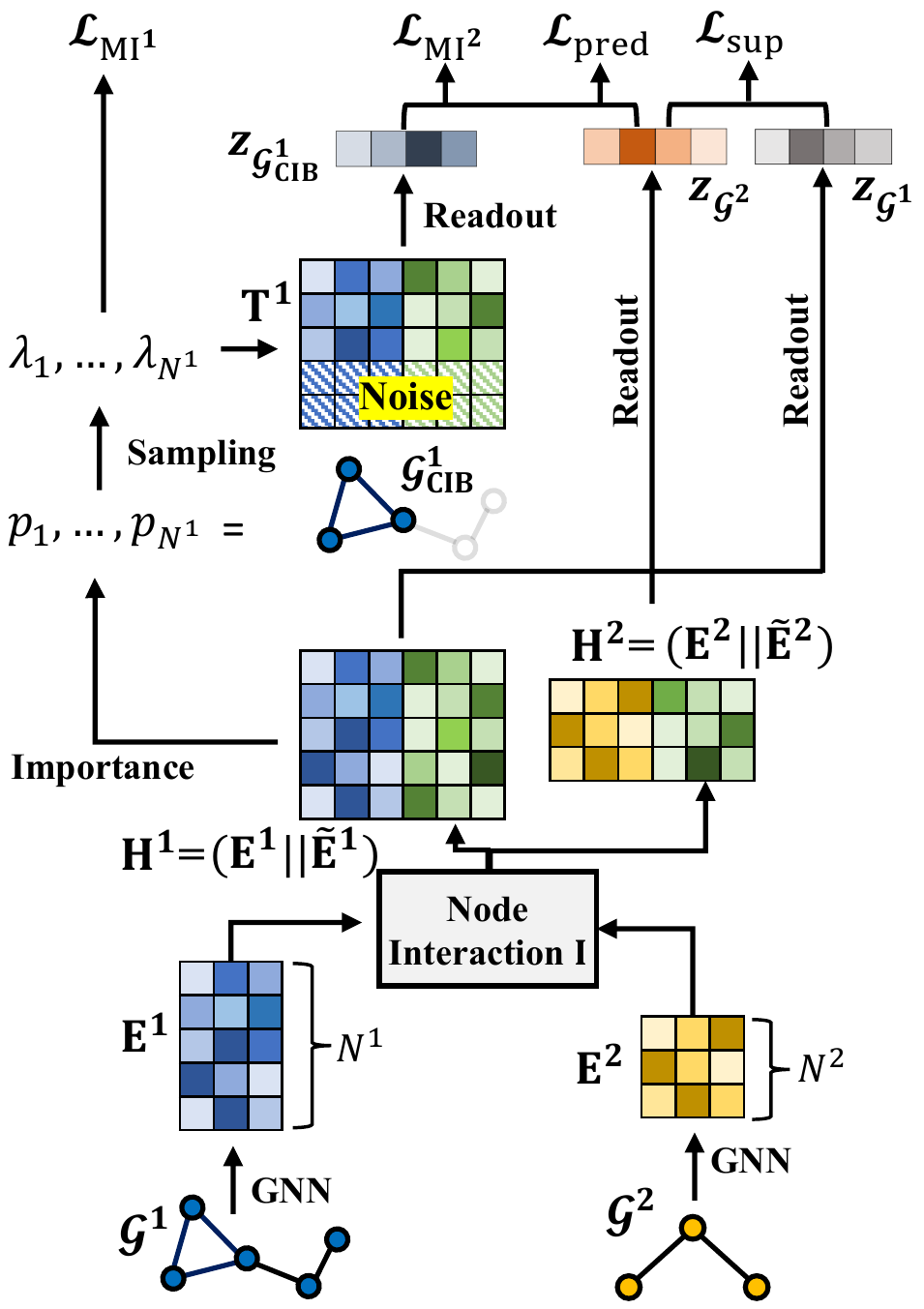}
    \caption{Overall model architecture.}
    \label{fig: Model Architecture}
\end{figure}

\subsection{Conditional Graph Information Bottleneck}
\label{Sec: Methodology Conditional Graph Information Bottleneck}
In this work, we are interested in learning the core subgraph $\mathcal{G}_{\mathrm{CIB}}^{1} = (\mathbf{X}_{\mathrm{CIB}}^{1}, \mathbf{A}_{\mathrm{CIB}}^{1})$ of input graph $\mathcal{G}^{1}$ conditioned on the paired input graph $\mathcal{G}^{2}$.

\smallskip
\begin{definition} 
\textit{(Conditional Information Bottleneck)}
Given random variables $X^{1}, X^{2}$, and $Y$, the Conditional Information Bottleneck (CIB) principle aims to compress $X^{1}$ to a bottleneck random variable $T^{1}$, while keeping the information relevant for predicting $Y$ conditioned on the random variable $X^{2}$:
\begin{equation}
\small
	\min_{T^{1}} -I(Y; T^{1} | X^{2}) + \beta I(X^{1}; T^{1} | X^{2})
\end{equation}
where $\beta$ is a Lagrangian multiplier for balancing the two conditional mutual information terms. In other words, $X^2$ is introduced as a condition in Equation \ref{eq: IB}.
\end{definition}

\smallskip
\begin{definition} 
\textit{(CIB-Graph)}
Given a pair of graphs $(\mathcal{G}^{1}, \mathcal{G}^{2})$ and its label information $\mathbf{Y}$, the optimal graph $\mathcal{G}_{\mathrm{CIB}}^{1} = (\mathbf{X}_{\mathrm{CIB}}^{1}, \mathbf{A}_{\mathrm{CIB}}^{1})$ discovered under the CIB principle is denoted as CIB-Graph:
\begin{equation}
\small
	\mathcal{G}_{\mathrm{CIB}}^{1} = \argmin_{\mathcal{G}_{\mathrm{CIB}}^{1}} -I(\mathbf{Y} ;\mathcal{G}_{\mathrm{CIB}}^{1} | \mathcal{G}^{2}) + \beta I( \mathcal{G}^{1} ; \mathcal{G}_{\mathrm{CIB}}^{1} | \mathcal{G}^{2})
\label{Eq: CIB Graph}
\end{equation}
where $\mathbf{X}_{\mathrm{CIB}}^{1}$ and $\mathbf{A}_{\mathrm{CIB}}^{1}$ denote the task-relevant feature and adjacency matrix of $\mathcal{G}^{1}$ conditioned on $\mathcal{G}^{2}$, respectively.
\end{definition}

It appears that the first term $-I(\mathbf{Y} ;\mathcal{G}_{\mathrm{CIB}}^{1} | \mathcal{G}^{2})$ is the prediction term, which encourages the optimal graph $\mathcal{G}_{\mathrm{CIB}}^{1}$ to capture sufficient information for predicting $\mathbf{Y}$ conditioned on the paired graph $\mathcal{G}^{2}$.
The second term $I(\mathcal{G}^{1}; \mathcal{G}_{\mathrm{CIB}}^{1} | \mathcal{G}^{2})$ is the compression term, which compresses $\mathcal{G}^{1}$ into $\mathcal{G}_{\mathrm{CIB}}^{1}$ conditioned on the paired graph $\mathcal{G}^{2}$.
{Consequently, jointly optimizing the two terms allows $\mathcal{G}_{\mathrm{CIB}}^{1}$ to preserve task relevant information of $\mathcal{G}^{1}$ conditioned on $\mathcal{G}^{2}$.} 
Next, to justify the model objective in Equation \ref{Eq: CIB Graph}, we introduce the following lemma.


\smallskip
\begin{lemma}
\label{lemma: nuisance invariance}
(Nuisance Invariance)
Given a pair of graphs $(\mathcal{G}^{1}, \mathcal{G}^{2})$ and its label information $\mathbf{Y}$, let $\mathcal{G}_{n}^{1}$ be a task irrelevant noise in the input graph $\mathcal{G}^{1}$.
Then, the following inequality holds:
\begin{equation}
\small
	I(\mathcal{G}_{\mathrm{CIB}}^{1}; \mathcal{G}_{n}^{1} | \mathcal{G}^{2}) \leq
	- I(\mathbf{Y}; \mathcal{G}_{\mathrm{CIB}}^{1} | \mathcal{G}^{2})
	+ I(\mathcal{G}^{1} ; \mathcal{G}_{\mathrm{CIB}}^{1} | \mathcal{G}^{2})
\end{equation}
\label{Lemma: Noise Invariance}
\vspace{-3ex}
\end{lemma}

Lemma \ref{Lemma: Noise Invariance} indicates that the \proposed~objective in Equation \ref{Eq: CIB Graph} is an upper bound of the conditional mutual information $I(\mathcal{G}_{\mathrm{CIB}}^{1}; \mathcal{G}_{n}^{1} | \mathcal{G}^{2})$ when $\beta = 1$.
That is, by optimizing Equation \ref{Eq: CIB Graph}, $\mathcal{G}_{\mathrm{CIB}}^{1}$ will be optimized to be less related to the task-irrelevant subgraph $\mathcal{G}_{n}^{1}$ conditioned on $\mathcal{G}^2$.
Please refer to Appendix \ref{App: Proofs Lemma} for a detailed proof of Lemma \ref{Lemma: Noise Invariance}.

\subsection{Model Architecture}
\label{Sec: Model Architecture}
We implement \proposed~based on the architecture of CIGIN \cite{CIGIN}, which is a simple and intuitive architecture designed for {molecular} relational learning.
Specifically, given a pair of graphs $\mathcal{G}^{1} = (\mathbf{X}^{1}, \mathbf{A}^{1})$ and $\mathcal{G}^{2} = (\mathbf{X}^{2}, \mathbf{A}^{2})$, we first generate a node embedding matrix for each graph with a GNN-based encoder as follows:
\begin{equation}
\small
    \mathbf{E}^{1} = \text{GNN}(\mathbf{X}^{1}, \mathbf{A}^{1}), \,\,\,\,
    \mathbf{E}^{2} = \text{GNN}(\mathbf{X}^{2}, \mathbf{A}^{2})
\end{equation}
where $\mathbf{E}^1\in\mathbb{R}^{N^1\times d}$ and $\mathbf{E}^2\in\mathbb{R}^{N^2\times d}$ are node embedding matrices for $\mathcal{G}^1$ and $\mathcal{G}^2$, respectively, and $N^1$ and $N^2$ denote the number of nodes in $\mathcal{G}^1$ and $\mathcal{G}^2$, respectively. 
Then, we model the node-wise interaction between $\mathcal{G}^{1}$ and $\mathcal{G}^{2}$ via an interaction map $\mathbf{I} \in \mathbb{R}^{N^{1} \times N^{2}}$ defined as  $\mathbf{I}_{ij} = \mathsf{sim}(\mathbf{E}_{i}^{1}, \mathbf{E}_{j}^{2})$, 
where $\mathsf{sim} (\cdot, \cdot)$ indicates the cosine similarity.
Then, we compute the embedding matrices $\tilde{\mathbf{E}}^{1} \in \mathbb{R}^{N^{1} \times d}$ and $\tilde{\mathbf{E}}^{2} \in \mathbb{R}^{N^{2} \times d}$ each of which regards its paired graph, based on the interaction map as $\tilde{\mathbf{E}}^{1} = \mathbf{I} \cdot \mathbf{E}^{2}$, and $\tilde{\mathbf{E}}^{2} = \mathbf{I}^\top \cdot \mathbf{E}^{1}$,
where $\cdot$ indicates matrix multiplication between two matrices.
Thus, $\tilde{\mathbf{E}}^{1}$ is the node embedding matrix of $\mathcal{G}^1$ that captures the interaction of nodes in $\mathcal{G}^1$ with those in $\mathcal{G}^2$, and likewise for $\tilde{\mathbf{E}}^{2}$.
Then, we generate the final node embedding matrix of $\mathcal{G}^1$, i.e., $\mathbf{H}^{1}$, by concatenating $\mathbf{E}^{1}$ and $\tilde{\mathbf{E}}^{1}$, i.e., $\mathbf{H}^{1} = (\mathbf{E}^{1} || \tilde{\mathbf{E}}^{1})\in \mathbb{R}^{N^1 \times 2d}$. The final node embedding matrix for $\mathcal{G}^2$, i.e., $\mathbf{H}^{2}$, is generated in a similar way.
{Lastly, we use Set2Set~\cite{vinyals2015order} as the graph readout function to generate the graph level embedding $\mathbf{z}_{\mathcal{G}^1}$ and $\mathbf{z}_{\mathcal{G}^2}$ for each graph $\mathcal{G}^1$ and $\mathcal{G}^2$, respectively.}
The overall model architecture is depicted in Figure \ref{fig: Model Architecture}.

\subsection{Model Optimization}
\label{Sec: Model Optimization}
To train the model while simultaneously detecting the core subgraph, we optimize the model with the objective function defined in Equation~\ref{Eq: CIB Graph} as follows:
\begin{equation}
	\min \underbrace{-I(\mathbf{Y} ;\mathcal{G}_{\mathrm{CIB}}^{1} | \mathcal{G}^{2})}_\text{Section \ref{sec:cond}} + \beta \underbrace{I( \mathcal{G}^{1};\mathcal{G}_{\mathrm{CIB}}^{1} | \mathcal{G}^{2})}_\text{Section \ref{sec:compression}} ,
\label{Eq: Model Objective}
\end{equation}
where each term indicates the prediction and compression, respectively.
In the following sections, we provide the upper bound of each term, which should be minimized during training.

\subsubsection{\textbf{Minimizing $-I(\mathbf{Y} ;\mathcal{G}_{\mathrm{CIB}}^{1} | \mathcal{G}^{2})$}}
\label{sec:cond}
Following the chain rule of mutual information, the first term $-I(\mathbf{Y} ;\mathcal{G}_{\mathrm{CIB}}^{1} | \mathcal{G}^{2})$, which aims to keep task-relevant information in $\mathcal{G}_{\mathrm{CIB}}^{1}$ conditioned on $\mathcal{G}^{2}$, can be decomposed as follows:
\begin{equation}
    -I(\mathbf{Y} ;\mathcal{G}_{\mathrm{CIB}}^{1} | \mathcal{G}^{2}) = -I(\mathbf{Y} ;\mathcal{G}_{\mathrm{CIB}}^{1}, \mathcal{G}^{2}) + I(\mathbf{Y} ; \mathcal{G}^{2}).
    \label{eqn:decompose}
\end{equation}
However, we empirically find out that minimizing $I(\mathbf{Y} ; \mathcal{G}^{2})$ deteriorates the model performance (See Appendix \ref{App: Sensitivity Analysis Gamma}).
Thus, we only consider $-I(\mathbf{Y} ;\mathcal{G}_{\mathrm{CIB}}^{1}, \mathcal{G}^{2})$ in this work.

\smallskip
\begin{proposition}
\label{Proposition: upperbound1}
(Upper bound of $-I(\mathbf{Y} ;\mathcal{G}_{\mathrm{CIB}}^{1}, \mathcal{G}^{2})$)
Given a pair of graph $(\mathcal{G}^{1}, \mathcal{G}^{2})$, its label information $\mathbf{Y}$, and the learned CIB-graph $\mathcal{G}_{CIB}^{1}$, we have
\begin{equation}
	-I(\mathbf{Y} ;\mathcal{G}_{\mathrm{CIB}}^{1}, \mathcal{G}^{2}) \leq \mathbb{E}_{\mathcal{G}_{\mathrm{CIB}}^{1}, \mathcal{G}^{2}, \mathbf{Y}}{[-\log{p_{\theta}(\mathbf{Y}|\mathcal{G}_{\mathrm{CIB}}^{1}, \mathcal{G}^{2})}]}
\end{equation}
\label{proposition: prediction}
where $p_{\theta}(\mathbf{Y}|\mathcal{G}_{\mathrm{CIB}}^{1}, \mathcal{G}^{2})$ is variational approximation of $p(\mathbf{Y}|\mathcal{G}_{\mathrm{CIB}}^{1}, \mathcal{G}^{2})$.
\end{proposition}
We model $p_{\theta}(\mathbf{Y}|\mathcal{G}_{\mathrm{CIB}}^{1}, \mathcal{G}^{2})$ as a predictor parametrized by $\theta$, which outputs the model prediction $\mathbf{Y}$ based on the input pair $(\mathcal{G}_{\mathrm{CIB}}^{1}, \mathcal{G}^{2})$.
Thus, we can minimize the upper bound of $-I(\mathbf{Y} ;\mathcal{G}_{\mathrm{\mathrm{CIB}}}^{1}, \mathcal{G}^{2})$ by minimizing the model prediction loss $\mathcal{L}_{\mathrm{pred}}(\mathbf{Y},\mathcal{G}_{\mathrm{CIB}}^{1}, \mathcal{G}^{2})$, which can be modeled as the cross entropy loss for classification and the mean square loss for regression.
A detailed proof for proposition \ref{proposition: prediction} is given in Appendix \ref{App: Proofs Proposition}.

\subsubsection{\textbf{Minimizing $I( \mathcal{G}^{1};\mathcal{G}_{\mathrm{CIB}}^{1} | \mathcal{G}^{2})$}}
\label{sec:compression}
For the second term of Equation~\ref{Eq: Model Objective}, i.e., $I( \mathcal{G}^{1};\mathcal{G}_{\mathrm{CIB}}^{1} | \mathcal{G}^{2})$, we decompose the term into the sum of two terms based on the chain rule of mutual information as follows:
\begin{equation}
I( \mathcal{G}^{1};\mathcal{G}_{\mathrm{CIB}}^{1} | \mathcal{G}^{2}) = I(\mathcal{G}_{\mathrm{CIB}}^{1}; \mathcal{G}^{1}, \mathcal{G}^{2}) - I(\mathcal{G}_{\mathrm{CIB}}^{1}; \mathcal{G}^{2}). 
\label{eqn:decomp2}
\end{equation}

Intuitively, minimizing $I(\mathcal{G}_{\mathrm{CIB}}^{1}; \mathcal{G}^{1}, \mathcal{G}^{2})$ aims to compress the information of a pair of graph $(\mathcal{G}^{1}, \mathcal{G}^{2})$ into $\mathcal{G}_{\mathrm{CIB}}^{1}$, while maximizing $I(\mathcal{G}_{\mathrm{CIB}}^{1}; \mathcal{G}^{2})$ encourages $\mathcal{G}_{\mathrm{CIB}}^{1}$ to keep the information about a paired graph $\mathcal{G}^{2}$ during compression. 

\smallskip
\noindent \underline{\textbf{Minimizing $I(\mathcal{G}_{\mathrm{CIB}}^{1}; \mathcal{G}^{1}, \mathcal{G}^{2})$}}.
Inspired by a recent approach on graph information bottleneck~\cite{VGIB}
that minimizes $I(\mathcal{G}_\mathrm{IB} ; \mathcal{G})$ by injecting noise into node representations,
we compress the information contained in $\mathcal{G}^{1}$ and $\mathcal{G}^{2}$ into $\mathcal{G}^{1}_\mathrm{CIB}$ by injecting noise into the learned node representation $\mathbf{H}^{1}$ that contains information regarding both $\mathcal{G}^{1}$ and $\mathcal{G}^{2}$.
The key idea is to enable the model to inject noise into insignificant subgraphs, while injecting less noise into more informative ones.
{More precisely, given a node $i$'s embedding $\mathbf{H}_{i}^{1}$, we calculate the probability $p_i$ with MLP, i.e., $p_i = \text{MLP}(\mathbf{H}_{i}^{1})$.}
With the calculated probability $p_i$, we replace the representation $\mathbf{H}_{i}^{1}$ of node $i$ with noise $\epsilon$, i.e., $\mathbf{T}_{i}^{1} = \lambda_{i}\mathbf{H}_{i}^{1}+ (1-\lambda_{i})\epsilon$, 
where $\lambda_{i} \sim \text{Bernoulli}(\text{Sigmoid}(p_i))$ and $\epsilon \sim N(\mu_{\mathbf{H}^{1}}, \sigma_{\mathbf{H}^{1}}^{2})$. 
Note that $\mu_{\mathbf{H}^{1}}$ and $\sigma_{\mathbf{H}^{1}}^{2}$ are mean and variance of $\mathbf{H}^{1}$, respectively.
Thus, the information of $\mathcal{G}^{1}$ and $\mathcal{G}^{2}$ are compressed into $\mathcal{G}_\text{CIB}^{1}$ with the probability of $p_{i}$ by replacing non-important nodes with noise.
That is, $p_{i}$ controls the information flow of $\mathcal{G}^{1}$ and $\mathcal{G}^{2}$ into $\mathcal{G}_\text{CIB}^{1}$.
Moreover, to make the sampling process differentiable, we adopt gumbel sigmoid \cite{maddison2016concrete,jang2016categorical} for discrete random variable $\lambda_{i}$, i.e., $\lambda_{i} = \text{Sigmoid}(1/t\log[{p_i/(1-p_i)}]+\log{[u/(1-u)]})$ where $u \sim \text{Uniform}(0, 1)$, and $t$ is the temperature hyperparameter.
Finally, we minimize the upper bound of $I(\mathcal{G}_{\mathrm{CIB}}^{1}; \mathcal{G}^{1}, \mathcal{G}^{2})$ as follows:
\begin{equation}
\begin{split}
	I(\mathcal{G}_{\mathrm{CIB}}^{1}; \mathcal{G}^{1}, \mathcal{G}^{2}) & \leq \mathbb{E}_{\mathcal{G}^{1}, \mathcal{G}^{2}}{[-\frac{1}{2} \log A + \frac{1}{2N^{1}}A + \frac{1}{2N^{1}} B^{2}]}
	\\
	& \coloneqq \mathcal{L}_{\mathrm{MI}^{1}}(\mathcal{G}_{\mathrm{CIB}}^{1}, \mathcal{G}^{1}, \mathcal{G}^{2})
\end{split}
\label{Eq: upper bound of GCIB1 G1 G2}
\end{equation}
where $A = \sum_{j = 1}^{N^{1}}{(1 - \lambda_{j})^{2}}$ and $B = \frac{\sum_{j = 1}^{N^{1}}{\lambda_{j}(\mathbf{H}_{j}^{1}-\mu_{\mathbf{H}^{1}})}}{\sigma_{\mathbf{H}^{1}}}$. 
A detailed proof for Equation \ref{Eq: upper bound of GCIB1 G1 G2} is given in Appendix \ref{App: proofs of upperbound}.

\smallskip
\noindent \underline{\textbf{Minimizing $-I(\mathcal{G}_{\mathrm{CIB}}^{1}; \mathcal{G}^{2})$}}. In this section, we introduce two different approaches for minimizing $-I(\mathcal{G}_{\mathrm{CIB}}^{1}; \mathcal{G}^{2})$.

\noindent\textbf{1) Variational IB-based approach. }
For the upper bound of $-I(\mathcal{G}_{\mathrm{CIB}}^{1}; \mathcal{G}^{2})$, we adopt the variational IB-based approach \cite{VIB} as follows:
\begin{equation}
\small
\begin{split}
	-I(\mathcal{G}_{\mathrm{CIB}}^{1}; \mathcal{G}^{2}) & \leq \mathbb{E}_{\mathcal{G}_{\mathrm{CIB}}^{1}, \mathcal{G}^{2}}[-\log p_{\xi}(\mathcal{G}^{2}|\mathcal{G}_{\mathrm{CIB}}^{1})] \\ & \coloneqq \mathcal{L}_{\mathrm{MI}^{2}}(\mathcal{G}_{\mathrm{CIB}}^{1}, \mathcal{G}^{2})
\end{split}
\label{eq: bound3}
\end{equation}
where $p_{\xi}(\mathcal{G}^{2}|\mathcal{G}_{\mathrm{CIB}}^{1})$ is the variational approximation of $p(\mathcal{G}^{2}|\mathcal{G}_{\mathrm{CIB}}^{1})$.
Although there exist various modeling choices for $p_{\xi}$ such as an MLP with non-linearity, we use a single-layered linear transformation without non-linearity. We argue that as more learnable parameters are involved in $p_{\xi}$, information that is useful for predicting $\mathcal{G}^2$ would be included in the parameters of $p_{\xi}$ rather than in the representation of $\mathcal{G}_{\mathrm{CIB}}^{1}$ itself, which incurs information loss as our goal is to obtain a high-quality representation of $\mathcal{G}_{\mathrm{CIB}}^{1}$. We indeed show in Appendix \ref{App: Additional Experiments 1} that a shallow $p_{\xi}$ is superior to a deep $p_{\xi}$.

\smallskip
\noindent\textbf{2) Contrastive learning-based approach. }
Besides, recently proposed contrastive learning \cite{tian2020contrastive,hjelm2018learning,you2020graph,velivckovic2018deep}, which learns to pull/push positive/negative samples in the representation space, has been theoretically proven to maximize the mutual information between positive pairs.
Thus, we additionally propose a variant of \proposed, called \proposedcont, which minimizes the term $-I(\mathcal{G}_{\mathrm{CIB}}^{1}; \mathcal{G}^{2})$ by minimizing the contrastive loss rather than the upper bound defined in Equation \ref{eq: bound3} as follows:
\begin{equation}
	\mathcal{L}_{\mathrm{MI}^{2}} = -\frac{1}{K} \sum_{i = 1}^{K} \log \frac{\exp(\mathsf{sim}(\mathbf{z}_{\mathcal{G}_{\mathrm{CIB}, i}^{1}}, \mathbf{z}_{\mathcal{G}_{i}^{2}})/\tau)}{\sum_{j = 1, j \neq i}^{K}{\exp(\mathsf{sim}(\mathbf{z}_{\mathcal{G}_{\mathrm{CIB}, i}^{1}}, \mathbf{z}_{\mathcal{G}_{j}^{2}})/\tau)}}
\end{equation}
where $K$ and $\tau$ indicate the number of paired graphs in a batch and the temperature hyperparameter, respectively.

We argue that preserving the information of $\mathcal{G}^{2}$ in $\mathcal{G}_{\mathrm{CIB}}^{1}$ by minimizing $-I(\mathcal{G}_{\mathrm{CIB}}^{1}; \mathcal{G}^{2})$ is the key to success of \proposed, which enables the conditional information compression of \proposed.
We later demonstrate its importance in Section \ref{sec: Ablation Studies}.

\smallskip
\noindent\textbf{Final Objectives. }
Finally, we train the model with the final objective given as follows:
\begin{equation}
    \mathcal{L}_{\mathrm{total}} = \mathcal{L}_{\mathrm{sup}} + {\mathcal{L}_{\mathrm{pred}}} + \beta ({\mathcal{L}_{\mathrm{MI}^{1}}} + {\mathcal{L}_{\mathrm{MI}^{2}}})
\label{eqn:finalloss}
\end{equation}
where $\beta$ controls the trade-off between prediction and compression.
Note that the supervised loss, i.e., $\mathcal{L}_{\mathrm{sup}}$, calculates the loss between the model prediction given the pair of input graphs, i.e., $(\mathcal{G}^1$, $\mathcal{G}^2)$, and the target response, i.e., $\mathbf{Y}$, without detecting the core subgraphs.
Moreover, in Appendix \ref{App: Selection G1 G2}, we conduct analyses on the selection of $\mathcal{G}^{1}$ and $\mathcal{G}^{2}$, and provide a guidance regarding how to decide which one of the graphs to use as $\mathcal{G}^{1}$.

\section{Experiments}
\subsection{Experimental Setup}

\noindent \textbf{Datasets.}
We use \textbf{eleven} datasets to comprehensively evaluate the performance of \proposed~ on three tasks, i.e., 1) molecular interaction prediction, 2) drug-drug interaction (DDI) prediction, and 3) graph similarity learning. 
Specifically, for the molecular interaction prediction task, we use Chromophore dataset \cite{Chromophore}, which is related to three optical properties of chromophores, as well as 5 other datasets, i.e., \textbf{MNSol} \cite{MNSol}, \textbf{FreeSolv} \cite{FreeSolv}, \textbf{CompSol} \cite{CompSol}, \textbf{Abraham} \cite{Abraham}, and \textbf{CombiSolv} \cite{CombiSolv}, which are related to the solvation free energy of solute.
In Chromophore dataset, maximum absorption wavelength (\textbf{Absorption}), maximum emission wavelength (\textbf{Emission}) and excited state lifetime (\textbf{Lifetime}) properties are used in this work.
For the DDI prediction task, we use 2 datasets, i.e., \textbf{ZhangDDI} \cite{ZhangDDI} and \textbf{ChChMiner} \cite{ChChMiner}, both of which contain labeled DDI data.
Lastly, for the graph similarity learning task, we use 3 datasets, i.e., \textbf{AIDS}, \textbf{IMDB} \cite{SimGNN}, and \textbf{OpenSSL} \cite{xu2017neural}, each of which is related to chemical compound, ego-network, and binary function's control flow, respectively. 
Further details on datasets are described in Appendix \ref{App: Datasets}.

\noindent \textbf{Methods Compared.}
For all three tasks, we compare~\proposed~with the state-of-the-art methods.
Specifically, for the molecular interaction prediction task, we mainly compare with CIGIN \cite{CIGIN}. 
For the DDI prediction task, we mainly compare with SSI-DDI \cite{nyamabo2021ssi} and MIRACLE \cite{MIRACLE}, and additionally compare with  CIGIN \cite{CIGIN} by changing its prediction head originally designed for regression to classification. 
Note that for the molecular interaction and DDI prediction tasks, we also compare~\proposed~with simple baseline methods, i.e., GCN \cite{GCN}, GAT \cite{GAT}, MPNN \cite{MPNN}, and GIN \cite{GIN}. Specifically, we independently encode a pair of graphs based on Set2Set pooling for fair comparisons, and concatenate the two encoded vectors to predict the target value by using an MLP.
Further details regarding the compared methods for the graph similarity learning task are described in Appendix \ref{App: Graph Similarity Learning}.

\begin{table*}[t]
    \centering
    \caption{Performance on molecular interaction prediction task (regression) in terms of RMSE.}
    \resizebox{0.95\linewidth}{!}{
    \begin{tabular}{lcccccccccc}
    \toprule
    \multirow{2}{*}{\textbf{Model}} & & \multicolumn{3}{c}{\textbf{Chromophore}} &  & \multirow{2}{*}{\textbf{MNSol}} & \multirow{2}{*}{\textbf{FreeSolv}} & \multirow{2}{*}{\textbf{CompSol}} & \multirow{2}{*}{\textbf{Abraham}} & \multirow{2}{*}{\textbf{CombiSolv}} \\ 
    \cmidrule{3-5}
    & & \textbf{Absorption}   & \textbf{Emission}     & \textbf{Lifetime}   &  & & & & & \\ 
    \midrule
    \multicolumn{11}{l}{\textbf{Interaction} \xmark} \\
    \midrule
    GCN     &    & $25.75 $ \scriptsize{(1.48)} & $31.87 $ \scriptsize{(1.70)} & $0.866 $ \scriptsize{(0.015)} &   & $0.675 $ \scriptsize{(0.021)} & $1.192 $ \scriptsize{(0.042)} & $0.389 $ \scriptsize{(0.009)} & $0.738 $ \scriptsize{(0.041)} & $0.672 $ \scriptsize{(0.022)} \\
    GAT     &    & $26.19 $ \scriptsize{(1.44)} & $30.90 $ \scriptsize{(1.01)} & $0.859 $ \scriptsize{(0.016)} &   & $0.731 $ \scriptsize{(0.007)} & $1.280 $ \scriptsize{(0.049)} & $0.387 $ \scriptsize{(0.010)} & $0.798 $ \scriptsize{(0.038)} &  $0.662 $ \scriptsize{(0.021)} \\
    MPNN    &    & $24.43 $ \scriptsize{(1.55)} & $30.17 $ \scriptsize{(0.99)} & $0.802 $ \scriptsize{(0.024)}  & & $0.682 $ \scriptsize{(0.017)} & $1.159 $ \scriptsize{(0.032)} & $0.359 $ \scriptsize{(0.011)} & $0.601 $ \scriptsize{(0.035)} & $0.568 $ \scriptsize{(0.005)} \\
    GIN     &    & $24.92 $ \scriptsize{(1.67)} & $32.31 $ \scriptsize{(0.26)} & $0.829 $ \scriptsize{(0.027)}  & & $0.669 $ \scriptsize{(0.017)} & $1.015 $ \scriptsize{(0.041)} & $0.331 $ \scriptsize{(0.016)} & $0.648 $ \scriptsize{(0.024)} & $0.595 $ \scriptsize{(0.014)} \\ 
    \midrule
    \multicolumn{11}{l}{\textbf{Interaction} \cmark} \\
    \midrule
    CIGIN     &  & $19.32 $ \scriptsize{(0.35)} & $25.09 $ \scriptsize{(0.32)} & $0.804 $ \scriptsize{(0.010)}  & & $0.607 $ \scriptsize{(0.024)} & $0.905 $ \scriptsize{(0.014)} & $0.308 $ \scriptsize{(0.018)} & $0.411 $ \scriptsize{(0.008)} & $0.451 $ \scriptsize{(0.009)} \\ 
    \midrule
    \proposed     &   & $\mathbf{17.87 }$ \scriptsize{(0.38)} & $24.44 $ \scriptsize{(0.21)} & $0.796 $ \scriptsize{(0.010)}  & & $0.568 $ \scriptsize{(0.013)} & $\mathbf{0.831 } $\scriptsize{(0.012)} & $0.277 $ \scriptsize{(0.008)} & $0.396 $ \scriptsize{(0.009)} & $0.428 $ \scriptsize{(0.009)} \\
    \proposedcont &  & $18.11 $ \scriptsize{(0.20)} & $\mathbf{23.90 }$ \scriptsize{(0.35)} & $\mathbf{0.771 }$ \scriptsize{(0.005)} &   & $\mathbf{0.538 }$ \scriptsize{(0.007)} & $0.852 $ \scriptsize{(0.022)} & $\mathbf{0.276 }$ \scriptsize{(0.017)} & $\mathbf{0.390 }$ \scriptsize{(0.006)} & $\mathbf{0.422 }$ \scriptsize{(0.005)} \\ 
    \bottomrule
    \end{tabular}}
    \label{tab: regression}
\end{table*}

\begin{table*}
    \centering
    \caption{Performance on drug-drug interaction prediction task (classification).}
    \resizebox{0.95\linewidth}{!}{
    
    \begin{tabular}{lcccccccccccc}
    \toprule
    \multirow{3}{*}{\textbf{Model}} & & \multicolumn{5}{c}{\textbf{(a) Transductive}} & & \multicolumn{5}{c}{\textbf{(b) Inductive}} \\ 
    \cmidrule{3-7} \cmidrule{9-13}
    & & \multicolumn{2}{c}{\textbf{ZhangDDI}} & & \multicolumn{2}{c}{\textbf{ChChMiner}} &  & \multicolumn{2}{c}{\textbf{ZhangDDI}} & & \multicolumn{2}{c}{\textbf{ChChMiner}} \\ 
    \cmidrule{3-4} \cmidrule{6-7} \cmidrule{9-10} \cmidrule{12-13}
    & & AUROC & Accuracy &  & AUROC & Accuracy & & AUROC & Accuracy & & AUROC & Accuracy \\ 
    \midrule
    \multicolumn{13}{l}{\textbf{Interaction} \xmark} \\
    \midrule
    GCN       &  & $91.64 $ \scriptsize{(0.31)} & $83.31 $ \scriptsize{(0.61)}&  & $94.71 $ \scriptsize{(0.33)} & $87.36 $ \scriptsize{(0.24)} & & $68.39 $ \scriptsize{(1.85)} & $63.78 $ \scriptsize{(1.55)} & & $73.63 $ \scriptsize{(0.44)} & $67.07 $ \scriptsize{(0.66)} \\
    GAT       &  & $92.10 $ \scriptsize{(0.28)} & $84.14 $ \scriptsize{(0.38)} & & $96.15 $ \scriptsize{(0.53)} & $89.49 $ \scriptsize{(0.88)} & & $69.99 $ \scriptsize{(2.95)} & $64.41 $ \scriptsize{(1.39)} & & $75.72 $ \scriptsize{(1.66)} & $68.77 $ \scriptsize{(1.48)} \\
    MPNN      &  & $92.34 $ \scriptsize{(0.35)} & $84.56 $ \scriptsize{(0.31)} & & $96.25 $ \scriptsize{(0.53)} & $90.02 $ \scriptsize{(0.42)} & & $71.54 $ \scriptsize{(1.24)} & $65.12 $ \scriptsize{(1.14)} &  & $75.45 $ \scriptsize{(0.32)} & $68.24 $ \scriptsize{(1.42)} \\
    GIN       &  & $93.16 $ \scriptsize{(0.04)} & $85.59 $ \scriptsize{(0.05)} & & $97.52 $ \scriptsize{(0.05)} & $91.89 $ \scriptsize{(0.66)} & & $72.74 $ \scriptsize{(1.32)} & $66.16 $ \scriptsize{(1.21)} &  & $74.63 $ \scriptsize{(0.48)} & $67.80 $ \scriptsize{(0.46)} \\ 
    \midrule
    \multicolumn{13}{l}{\textbf{Interaction} \cmark} \\
    \midrule
    SSI-DDI   &  & $92.74 $ \scriptsize{(0.12)} & $84.61 $ \scriptsize{(0.18)} & & $98.44 $ \scriptsize{(0.08)} & $93.50 $ \scriptsize{(0.16)} & & $73.29 $ \scriptsize{(2.23)} & $66.53 $ \scriptsize{(1.31)} &  & $78.24 $ \scriptsize{(1.29)} & $70.69 $ \scriptsize{(1.47)}\\
    MIRACLE   &  & $93.05 $ \scriptsize{(0.07)} & $84.90 $ \scriptsize{(0.36)} & & $88.66 $ \scriptsize{(0.37)} & $84.29 $ \scriptsize{(0.14)} & & $73.23 $ \scriptsize{(3.32)} & $50.00 $ \scriptsize{(0.00)} &  & $60.25 $ \scriptsize{(0.56)} & $50.09 $ \scriptsize{(0.11)} \\ 
    CIGIN     &  & $93.28 $ \scriptsize{(0.13)} & $85.54 $ \scriptsize{(0.30)} & & $98.51 $ \scriptsize{(0.10)} & $93.77 $ \scriptsize{(0.25)} & & $74.02 $ \scriptsize{(0.10)} & $66.81 $ \scriptsize{(0.09)} &  & $79.23 $ \scriptsize{(0.51)} & $71.56 $ \scriptsize{(0.38)} \\ 
    \midrule
    \proposed      &  & $\mathbf{94.74 }$ \scriptsize{(0.47)} & $\mathbf{86.88 }$ \scriptsize{(0.56)} &  & $98.80 $ \scriptsize{(0.04)} & $\mathbf{94.69 }$ \scriptsize{(0.16)} & & $74.59 $ \scriptsize{(0.88)} & $\mathbf{67.65 }$ \scriptsize{(1.07)} &  & $81.14$ \scriptsize{(1.20)} & $72.47$ \scriptsize{(0.16)}  \\
    \proposedcont &  & $93.78 $ \scriptsize{(0.62)} & $86.36 $ \scriptsize{(0.75)} &  & $\mathbf{98.84 }$ \scriptsize{(0.31)} & $94.52 $ \scriptsize{(0.38)} & & $\mathbf{75.08 }$ \scriptsize{(0.34)} & $67.31 $ \scriptsize{(0.82)} &  & $\mathbf{81.51} $ \scriptsize{(0.67)} & $\mathbf{74.29} $ \scriptsize{(0.14)} \\ 
    \bottomrule
    \end{tabular}}
    \label{tab: classification}
\end{table*}

\noindent \textbf{Evaluation Metrics.}
The performance of the molecular interaction prediction task is evaluated in terms of RMSE~\cite{CIGIN}, that of the drug-drug interaction prediction task is evaluated in terms of AUROC and accuracy~\cite{MIRACLE}, and that of similarity learning task is evaluated in terms of MSE, Spearman's Rank Correlation Coefficient (denoted as $\rho$), and precision@10 (p@10)~\cite{H2MN}.
{We further provide the details on the evaluation protocol in Appendix \ref{App: Evaluation Protocol}.}

\noindent \textbf{Implementation Details.}
For the molecular interaction prediction and graph similarity learning tasks, we use 3-layer MPNN \cite{MPNN} and GCN \cite{GCN} as our backbone graph encoder, respectively, and a 3-layer MLP with ReLU activation as the predictor following the previous works \cite{CIGIN,H2MN}. 
For the drug-drug interaction prediction task, we use GIN \cite{GIN} as our backbone graph encoder, and a single layer MLP without activation as the predictor.
Hyperparameter details are described in Appendix \ref{App: Implementation Details}.

\begin{figure*}[t]
    \centering
    \includegraphics[width=1.0\linewidth]{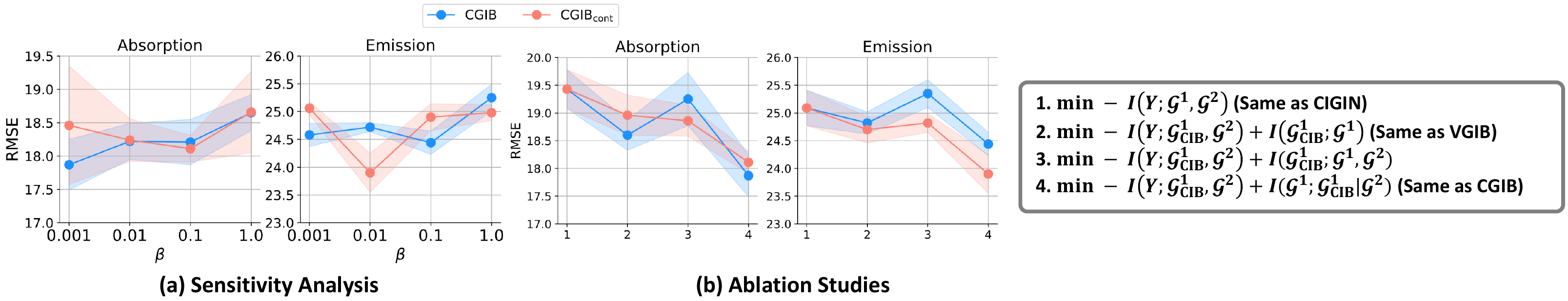}
    \caption{Model analysis.}
    \label{fig: Model analysis}
\end{figure*}

\subsection{Overall Performance}
\label{sec: Overall Performance}
The empirical performance of \proposed~on molecular interaction prediction and drug-drug interaction prediction tasks
is summarized in Table \ref{tab: regression} and Table \ref{tab: classification}, respectively.
We have the following observations:
\textbf{1)} \proposed~outperforms all other baseline methods that overlook the significance of the core subgraph during training in both molecular interaction prediction task (i.e., CIGIN), and drug-drug interaction prediction task (SSI-DDI and MIRACLE).
We argue that \proposed~improves its generalization ability by making predictions based on the detected core subgraph of the given graph, which is the minimal structure sufficient to represent the properties of the graph.
Considering that a certain functional group induces the same or similar chemical reactions, learning from the core substructure is crucial, especially in chemical reaction prediction tasks.
\textbf{2)} To further demonstrate the generalization ability of \proposed, we conduct additional experiments in the inductive setting (Table \ref{tab: classification}(b)), which is more practical and closer to the real-world applications. We observe that~\proposed~consistently outperforms other baseline methods in the inductive setting as well, which verifies the practicality of~\proposed.
We argue that as~\proposed~makes predictions based on the core subgraphs of graphs that are shared among the total set of graphs, i.e., $\mathbb{G}$ (Refer to Appendix \ref{App: Evaluation Protocol}),~\proposed~can make accurate predictions even though the graphs in the test set are not seen during training.
Based on the results of these two experiments, we argue that the key to the success of \proposed~is the generalization ability thanks to the detection of the core subgraph during training.
\textbf{3)} It is worth noting that simple baseline methods that naively concatenate the representations of a pair of graphs, i.e., GCN, GAT, MPNN, and GIN, generally perform worse than the methods that consider the interaction between the graphs, i.e., CIGIN, SSI-DDI, and MIRACLE, which implies that modeling the interaction between graphs is important in relational learning framework.
To verify the wide applicability of \proposed, we conduct experiments on the graph similarity learning task in Appendix \ref{App: Graph Similarity Learning}.



\subsection{Model Analysis}
\label{sec: Model Analysis}

\subsubsection{Sensitivity Analysis on $\beta$}
\label{sec: Sensitivity Analysis}

In this section, we analyze the effect of $\beta$, which controls the trade-off between the prediction and compression in our final objectives shown in Equation~\ref{eqn:finalloss}.
As shown in Figure \ref{fig: Model analysis} (a), there exists the optimal point of $\beta$ in terms of the model performance, which indicates the existence of the trade-off between the prediction and the information compression. 
We have the following observations:
\textbf{1)} The model consistently performs the worst when $\beta = 1$.
This is because $\beta = 1$ encourages the model to aggressively compress the information of the input graph, thereby hardly capturing the core subgraph that is related to the target task.
\textbf{2)} On the other hand, decreasing $\beta$ does not always lead to a good performance.
{Recall that decreasing $\beta$ encourages the model to keep the original information of the given graph structure.}
In an extreme case, i.e., when $\beta = 0$, the model would not consider the information compression at all.
In this case, since the model only needs to focus on the prediction term, prediction can be made using the entire graph structure without finding the core subgraph, which consequently leads to the lack of the generalization ability.
Recent works on augmentation \cite{gontijo2020tradeoffs,zhu2021graph} can also shed light on this phenomenon, i.e., $\beta$ controls the trade-off between affinity and diversity.
We also conduct qualitative analysis on $\beta$ in Appendix \ref{App: Additional Experiments 2}.


\subsubsection{{Ablation Studies}}
\label{sec: Ablation Studies}
To verify the benefit of the conditional compression module of \proposed, i.e., ${I( \mathcal{G}^{1};\mathcal{G}_{\mathrm{CIB}}^{1} | \mathcal{G}^{2})}$ described in Section~\ref{sec:compression}, we conduct ablation studies on two datasets, i.e., Absorption and Emission, in Figure \ref{fig: Model analysis} (b).
Recall that the conditional mutual information $I(\mathcal{G}^{1}; \mathcal{G}^{1}_{\mathrm{CIB}} | \mathcal{G}^{2})$ is decomposed as shown in Equation~\ref{eqn:decomp2}: $I(\mathcal{G}^{1}; \mathcal{G}^{1}_{\mathrm{CIB}} | \mathcal{G}^{2}) = I(\mathcal{G}^{1}_{\mathrm{CIB}}; \mathcal{G}^{1}, \mathcal{G}^{2}) - I(\mathcal{G}^{1}_{\mathrm{CIB}}; \mathcal{G}^{2})$.
We have the following observations:
\textbf{1)} Existing methods that only take a single graph into account (i.e., $I(\mathcal{G}^{1}_{\mathrm{CIB}}; \mathcal{G}^{1})$), also perform better than the baseline that does not consider the subgraph at all (i.e., Without IB).
This implies that considering the core subgraph of the given graph generally improves the model performance.
\textbf{2)} On the other hand, given two graphs $\mathcal{G}^{1}$ and $\mathcal{G}^{2}$, compressing the information by minimizing the conditional mutual information (i.e., $I(\mathcal{G}^{1}; \mathcal{G}^{1}_{\mathrm{CIB}} | \mathcal{G}^{2})$) instead of the joint mutual information (i.e., $I(\mathcal{G}^{1}_{\mathrm{CIB}}; \mathcal{G}^{1}, \mathcal{G}^{2})$) is crucial in relational learning.
This is because the conditional mutual information encourages the compressed subgraph $\mathcal{G}^{1}_{\mathrm{CIB}}$ of graph $\mathcal{G}^{1}$ to keep the information regarding $\mathcal{G}^{2}$, due to the additional $I(\mathcal{G}_\text{CIB}^{1}; \mathcal{G}^{2})$ maximization term that appears in the decomposition of $I(\mathcal{G}^{1}; \mathcal{G}^{1}_{\mathrm{CIB}} | \mathcal{G}^{2})$.
That is, by assuring $\mathcal{G}^{1}_{\mathrm{CIB}}$ to have information about $\mathcal{G}^{2}$, the model considers $\mathcal{G}^{2}$ during the compression procedure, which aligns with our conditional information bottleneck objective.
\textbf{3)} Moreover, compressing the information by minimizing the conditional mutual information consistently outperforms compressing the information solely based on a single graph, i.e., $I(\mathcal{G}^{1}_{\mathrm{CIB}}; \mathcal{G}^{1})$, {which is equivalent to VGIB \cite{VGIB}.}
This is because the core subgraph of $\mathcal{G}^{1}$ should be determined based on the other graph to be interacted with, i.e., $\mathcal{G}^{2}$.
{Therefore, we argue that the current graph information bottleneck approaches such as VGIB~\cite{VGIB}, GIB~\cite{PGIB}, and GSAT~\cite{GSAT}, are not suitable for relational learning tasks.}
\textbf{4)} What's interesting is that naively modeling the joint compression simply through the joint mutual information (i.e., $I(\mathcal{G}^{1}_{\mathrm{CIB}}; \mathcal{G}^{1}, \mathcal{G}^{2})$) performs even  worse than considering only single graph (i.e., $I(\mathcal{G}^{1}_{\mathrm{CIB}}; \mathcal{G}^{1})$).
{
This implies that if the compressed subgraph $\mathcal{G}^{1}_{\mathrm{CIB}}$ does not fully contain the information of $\mathcal{G}^{2}$, the information of $\mathcal{G}^{2}$ can interfere with the optimal compression of the pair ($\mathcal{G}^{1}$, $\mathcal{G}^{2}$) into $\mathcal{G}^{1}_{\mathrm{CIB}}$.
Due to the suboptimality, the model trained with the joint mutual information sometimes performs worse than the one that does not consider the subgraph at all (i.e., Without IB).
}

To summarize our findings, the ablation studies demonstrate that simply adopting IB principle into relational learning framework is not trivial, and that~\proposed~successfully adopts the IB principle for relational learning.

\begin{figure*}[t]
    \centering
    \includegraphics[width=0.8\linewidth]{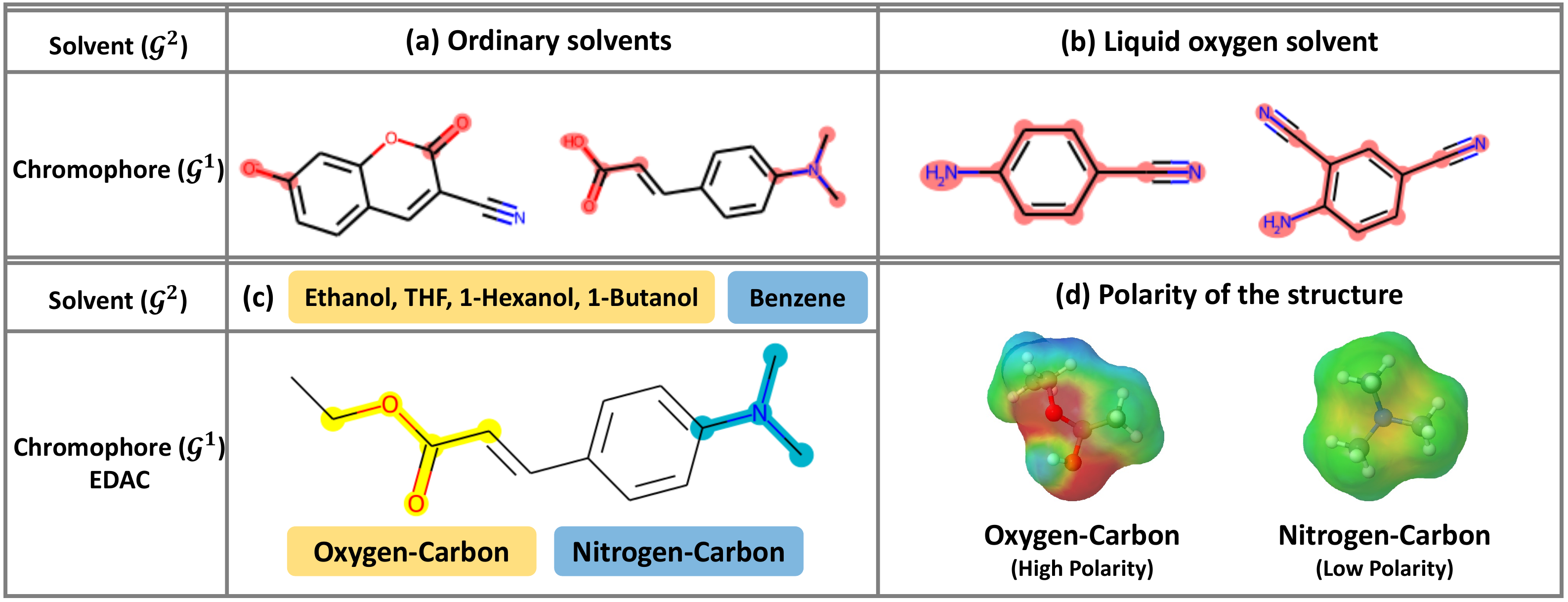}
    \caption{Qualitative analysis on CIB-Graph.}
    \label{fig: Qualitative Analysis}
\end{figure*}

\subsection{Qualitative Analysis on CIB-Graph}
We qualitatively analyze the substructures based on our prior chemical knowledge. 
In the Chromophore dataset, \proposed~predicts that the edge substructures of chromophores are important in the chromophore-solvent reactions, as shown in Figure \ref{fig: Qualitative Analysis}(a).
This prediction result of \proposed~ aligns with the chemical knowledge that chemical reactions usually happen around the ionized atoms \cite{hynes1985chemical}. 
Figure \ref{fig: Qualitative Analysis}(b) shows the important substructures of chromophores predicted by \proposed~when they react with liquid oxygen solvents.
As shown in the results, \proposed~predicts that the entire structure is important in the chemical reactions between chromophores and oxygen solvents. 
This result again aligns with the chemical knowledge that chemical reactions can happen in the entire molecule because the size of the oxygen solvent is small enough to permeate the chromophores.

We also find out that the important substructure of chromophores predicted by~\proposed~varies according to which solvent the chromophores react with.
Figure \ref{fig: Qualitative Analysis}(c) shows the important substructure in a chromophore named trans-ethyl p-(dimethylamino) cinamate (EDAC) \cite{singh2009effect} detected by \proposed~in five different solvents: benzene, ethanol, THF, 1-hexanol, and 1-butanol. 
We observe that \proposed~predicts that the nitrogen-carbon substructure (marked in blue) is important in the benzene solvent, whereas the oxygen-carbon substructure (marked in yellow) is predicted to be important in the ethanol, THF, 1-hexanol, and 1-butanol solvents. 
These results can be understood by the chemical polarity and the solvent solubility. 
Since nonpolar molecules usually interact with nonpolar molecules \cite{reichardt1965empirical}, the low-polarity nitrogen-carbon substructure is considered as an important substructure in benzene, which is a completely nonpolar solvent. 
On the other hand, polar molecules interact with polar molecules. 
Hence, the oxygen-carbon substructure with polarity is considered as an important substructure in the polar solvents ethanol and THF, all of which are polar solvents.
Note that the polarity of nitrogen-carbon substructure and oxygen-carbon structure is denoted in Figure \ref{fig: Qualitative Analysis} (d) \footnote{As colors differ in a molecule, its polarity gets higher.} left and right, respectively. 
Although 1-hexanol and 1-butanol are categorized into nonpolar solvents due to their overall weak polarity, the OH substructures in 1-hexanol and 1-butanol have a local polarity. 
For this reason, we conjecture that the oxygen-carbon substructure is predicted to be important in 1-hexanol and 1-butanol solvents.
We further provide a quantitative analysis on the selected CIB-Graph, i.e., $\mathcal{G}^1_\text{CIB}$ in Appendix \ref{App: Quantitative Analysis on CIB-Graph}.

\section{Conclusion}
In this paper, we propose a novel molecular relational learning framework, named \proposed, which predicts the interaction behavior between a pair of molecules by detecting important subgraphs therein.
The main idea is, given a pair of molecules, to find a substructure of a given molecule containing the minimal sufficient information regarding the task at hand conditioned on the paired molecule based on the principle of conditional graph information bottleneck.
By doing so, \proposed~adaptively selects the core substructure of the input molecule according to its paired molecule, which aligns with the nature of chemical reactions.
Our extensive experiments demonstrate that \proposed~consistently outperforms existing state-of-the-art methods in molecular relational learning tasks.
Moreover,~\proposed~provides convincing explanations regarding chemical reactions, which verifies its practicality in real-world applications.

\noindent\textbf{Acknowledgement}
This work was supported by Institute of Information \& communications Technology Planning \& Evaluation (IITP) grant funded by the Korea government(MSIT) (No.2022-0-00077), and core KRICT project from the Korea Research Institute of Chemical Technology (KK2351-10).
\nocite{langley00}

\clearpage

\bibliography{example_paper}
\bibliographystyle{icml2023}

\newpage
\appendix
\onecolumn

\section{Proofs}
\subsection{Proof of Lemma \ref{lemma: nuisance invariance}}
\label{App: Proofs Lemma}
Assuming that $\mathcal{G}^1, \mathcal{G}_{\mathrm{CIB}}^1, \mathcal{G}_{n}^1, \mathcal{G}^2,$ and $\mathbf{Y}$ satisfy the markov condition $(\mathbf{Y}, \mathcal{G}_{n}^1, \mathcal{G}^2) \rightarrow \mathcal{G}^1 \rightarrow \mathcal{G}_{\mathrm{CIB}}^1$, we have the following inequality due to data processing inequality:
\begin{equation}
\small
\begin{split}
    I(\mathcal{G}^{1} ; \mathcal{G}_{\mathrm{CIB}}^{1} | \mathcal{G}^{2}) &= I(\mathcal{G}_{\mathrm{CIB}}^{1}; \mathcal{G}^{1} , \mathcal{G}^{2}) - I(\mathcal{G}_{\mathrm{CIB}}^{1}; \mathcal{G}^{2})\\
    &\geq I(\mathcal{G}_{\mathrm{CIB}}^{1}; \mathbf{Y}, \mathcal{G}_{n}^{1} , \mathcal{G}^{2}) - I(\mathcal{G}_{\mathrm{CIB}}^{1}; \mathcal{G}^{2}) \\
    &= I(\mathcal{G}_{\mathrm{CIB}}^{1}; \mathcal{G}_{n}^{1} , \mathcal{G}^{2}) + I(\mathcal{G}_{\mathrm{CIB}}^{1}; \mathbf{Y} | \mathcal{G}_{n}^{1} , \mathcal{G}^{2}) - I(\mathcal{G}_{\mathrm{CIB}}^{1}; \mathcal{G}^{2}) \\
    &= I(\mathcal{G}_{\mathrm{CIB}}^{1}; \mathcal{G}_{n}^{1} | \mathcal{G}^{2}) + I(\mathcal{G}_{\mathrm{CIB}}^{1}; \mathbf{Y} | \mathcal{G}_{n}^{1} , \mathcal{G}^{2})
\end{split}
\raisetag{10pt}
\label{Eq: 17}
\end{equation}
Suppose that $\mathcal{G}_{n}^{1}$ and $\mathbf{Y}$, $\mathcal{G}_{n}^{1}$ and $\mathcal{G}^{2}$, and joint random variable $(\mathcal{G}_{n}^{1} , \mathcal{G}^{2})$ and $\mathbf{Y}$ are independent, respectively. 
Then, for $I(\mathcal{G}_{\mathrm{CIB}}^{1}; \mathbf{Y} | \mathcal{G}_{n}^{1} , \mathcal{G}^{2})$, we have:
\begin{equation}
\small
\begin{split}
    I(\mathcal{G}_{\mathrm{CIB}}^{1}; \mathbf{Y} | \mathcal{G}_{n}^{1} , \mathcal{G}^{2}) &= H(\mathbf{Y}| \mathcal{G}_{n}^{1}, \mathcal{G}^{2}) - H(\mathbf{Y}| \mathcal{G}_{n}^{1}, \mathcal{G}_{\mathrm{CIB}}^{1}, \mathcal{G}^{2})\\
    &\geq H(\mathbf{Y}| \mathcal{G}^{2}) - H(\mathbf{Y}| \mathcal{G}_{\mathrm{CIB}}^{1}, \mathcal{G}^{2}) \\
    &= I(\mathbf{Y}; \mathcal{G}_{\mathrm{CIB}}^{1}| \mathcal{G}^{2})
\end{split}
\label{Eq: 18}
\end{equation}
By plugging Equation~\ref{Eq: 18} into Equation~\ref{Eq: 17}, we have:
\begin{equation}
\small
    I(\mathcal{G}^{1}; \mathcal{G}_{\mathrm{CIB}}^{1} | \mathcal{G}^{2}) \geq I(\mathcal{G}_{\mathrm{CIB}}^{1}; \mathcal{G}_{n}^{1} | \mathcal{G}^{2}) + I(\mathbf{Y}; \mathcal{G}_{\mathrm{CIB}}^{1}| \mathcal{G}^{2})
\end{equation}

\subsection{Proof of Proposition \ref{Proposition: upperbound1}}
\label{App: Proofs Proposition}
By the definition of mutual information and introducing variational approximation $p_{\theta}(\mathbf{Y}|\mathcal{G}_{CIB}^{1}, \mathcal{G}^{2})$ of intractable distribution $p(\mathbf{Y}|\mathcal{G}_{CIB}^{1}, \mathcal{G}^{2})$, we have:
\begin{equation}
\small
\begin{split}
    I(\mathbf{Y} ; \mathcal{G}_{\mathrm{CIB}}^{1}, \mathcal{G}^{2}) &= \mathbb{E}_{\mathbf{Y}, \mathcal{G}_{\mathrm{CIB}}^{1}, \mathcal{G}^{2}}[\log \frac{p(\mathbf{Y}|\mathcal{G}_{\mathrm{CIB}}^{1}, \mathcal{G}^{2})}{p(\mathbf{Y})}]\\
    &= \mathbb{E}_{\mathbf{Y}, \mathcal{G}_{\mathrm{CIB}}^{1}, \mathcal{G}^{2}}[\log \frac{p_{\theta}(\mathbf{Y}|\mathcal{G}_{\mathrm{CIB}}^{1}, \mathcal{G}^{2})}{p(\mathbf{Y})}] \\
    &+ \mathbb{E}_{\mathcal{G}_{\mathrm{CIB}}^{1}, \mathcal{G}^{2}}[KL(p(\mathbf{Y}|\mathcal{G}_{\mathrm{CIB}}^{1}, \mathcal{G}^{2}) || p_{\theta}(\mathbf{Y}|\mathcal{G}_{\mathrm{CIB}}^{1}, \mathcal{G}^{2}))]\\
\end{split}
\raisetag{30pt}
\end{equation}
According to the non-negativity of the KL divergence, we have:
\begin{equation}
\small
\begin{split}
    I(\mathbf{Y} ; \mathcal{G}_{\mathrm{CIB}}^{1}, \mathcal{G}^{2}) & \geq \mathbb{E}_{\mathbf{Y}, \mathcal{G}_{\mathrm{CIB}}^{1}, \mathcal{G}^{2}}[\log \frac{p_{\theta}(\mathbf{Y}|\mathcal{G}_{\mathrm{CIB}}^{1}, \mathcal{G}^{2})}{p(\mathbf{Y})}] \\
    &= \mathbb{E}_{\mathbf{Y}, \mathcal{G}_{\mathrm{CIB}}^{1}, \mathcal{G}^{2}}[\log p_{\theta}(\mathbf{Y}|\mathcal{G}_{\mathrm{CIB}}^{1}, \mathcal{G}^{2})] + H(\mathbf{Y}) \\
\end{split}
\end{equation}

\subsection{Proof of Equation \ref{Eq: upper bound of GCIB1 G1 G2}}
\label{App: proofs of upperbound}
Given the perturbed graph $\mathcal{G}_{\mathrm{CIB}}^{1}$ and its representation $\mathbf{z}_{\mathcal{G}_{\mathrm{CIB}}^{1}}$, we assume there is no information loss during the readout process, i.e., $I(\mathbf{z}_{\mathcal{G}_{\mathrm{CIB}}^{1}}; \mathcal{G}^{1}, \mathcal{G}^{2}) \approx I(\mathcal{G}_{\mathrm{CIB}}^{1}; \mathcal{G}^{1}, \mathcal{G}^{2})$.
Now, we derive the upper bound of $I(\mathbf{z}_{\mathcal{G}_{\mathrm{CIB}}^{1}}; \mathcal{G}^{1}, \mathcal{G}^{2})$ by introducing the variation approximation $q(\mathbf{z}_{\mathcal{G}_{\mathrm{CIB}}^{1}})$ of distribution $p(\mathbf{z}_{\mathcal{G}_{\mathrm{CIB}}^{1}})$:
\begin{equation}
\small
\begin{split}
    I(\mathbf{z}_{\mathcal{G}_{\mathrm{CIB}}^{1}}; \mathcal{G}^{1}, \mathcal{G}^{2}) &= \mathbb{E}_{\mathbf{z}_{\mathcal{G}_{\mathrm{CIB}}^{1}}, \mathcal{G}^{1}, \mathcal{G}^{2}}[\log \frac{p_{\phi}(\mathbf{z}_{\mathcal{G}_{\mathrm{CIB}}^{1}}|\mathcal{G}^{1}, \mathcal{G}^{2})}{p(\mathbf{z}_{\mathcal{G}_{\mathrm{CIB}}^{1}})}]\\
    &= \mathbb{E}_{\mathcal{G}^{1}, \mathcal{G}^{2}}[\log \frac{p_{\phi}(\mathbf{z}_{\mathcal{G}_{\mathrm{CIB}}^{1}}|\mathcal{G}^{1}, \mathcal{G}^{2})}{q(\mathbf{z}_{\mathcal{G}_{\mathrm{CIB}}^{1}})}] \\
    &- \mathbb{E}_{\mathbf{z}_{\mathcal{G}_{\mathrm{CIB}}^{1}}, \mathcal{G}^{1}, \mathcal{G}^{2}}[KL ( p(\mathbf{z}_{\mathcal{G}_{\mathrm{CIB}}^{1}}) || q(\mathbf{z}_{\mathcal{G}_{\mathrm{CIB}}^{1}}))]\\
\end{split}
\end{equation}
According to the non-negativity of KL divergence, we have:
\begin{equation}
\small
    I(\mathbf{z}_{\mathcal{G}_{\mathrm{CIB}}^{1}}; \mathcal{G}^{1}, \mathcal{G}^{2}) \leq \mathbb{E}_{\mathcal{G}^{1}, \mathcal{G}^{2}}[KL(p_{\phi}(\mathbf{z}_{\mathcal{G}_{\mathrm{CIB}}^{1}}|\mathcal{G}^{1}, \mathcal{G}^{2}) || q(\mathbf{z}_{\mathcal{G}_{\mathrm{CIB}}^{1}}))]
\label{Eq: bound2}
\raisetag{20pt}
\end{equation}
Following VIB \cite{VIB}, we assume that $q(\mathbf{z}_{\mathcal{G}_{\mathrm{CIB}}^{1}})$ is obtained by aggregating the node representations in a fully perturbed graph.
The noise $\epsilon \sim N(\mu_{\mathbf{H}^{1}}, \sigma_{\mathbf{H}^{1}}^{2})$ is sampled from a Gaussian distribution where $\mu_{\mathbf{H}^{1}}$ and $\sigma_{\mathbf{H}^{1}}^{2}$ are mean and variance of $\mathbf{H}^{1}$ which contains information of both $\mathcal{G}^{1}$ and $\mathcal{G}^{2}$.
Choosing sum pooling as the readout function, since the summation of Gaussian distributions is a Gaussian, we have the following equation:
\begin{equation}
\small
    q(\mathbf{z}_{\mathcal{G}_{\mathrm{CIB}}^{1}}) = \mathcal{N}(N^{1} \mu_{\mathbf{H}^{1}}, N^{1} \sigma_{\mathbf{H}^{1}}^{2})
\label{Eq: variational q}
\end{equation}
Then for $p_{\phi}(\mathbf{z}_{\mathcal{G}_{\mathrm{CIB}}^{1}}| \mathcal{G}^{1}, \mathcal{G}^{2})$, we have the following equation:
\begin{equation}
\small
    p_{\phi}(\mathbf{z}_{\mathcal{G}_{\mathrm{CIB}}^{1}}|\mathcal{G}^{1}, \mathcal{G}^{2}) = \mathcal{N}(N^{1} \mu_{\mathbf{H}^{1}} + \sum_{j = 1}^{N^{1}}{\lambda_{j} \mathbf{H}_{j}^{1}} - \sum_{j = 1}^{N^{1}}{\lambda_{j} \mu_{\mathbf{H}^{1}}} , \sum_{j = 1}^{N^{1}}{(1 - \lambda_{j})^{2}\sigma_{\mathbf{H}^{1}}^{2}})
    \raisetag{10pt}
\label{Eq: original p}
\end{equation}
Finally, we have following inequality by plugging Equation \ref{Eq: variational q} and Equation \ref{Eq: original p} into Equation Equation \ref{Eq: bound2}:
\begin{equation}
\small
    I(\mathbf{z}_{\mathcal{G}_{\mathrm{CIB}}^{1}}; \mathcal{G}^{1}, \mathcal{G}^{2}) \leq \mathbb{E}_{\mathcal{G}^{1}, \mathcal{G}^{2}}{[-\frac{1}{2} \log A + \frac{1}{2N^{1}}A + \frac{1}{2N^{1}} B^{2}]} + C
\end{equation}
where $A = \sum_{j = 1}^{N^{1}}{(1 - \lambda_{j})^{2}}$, $B = \frac{\sum_{j = 1}^{N^{1}}{\lambda_{j}(\mathbf{H}_{j}^{1}-\mu_{\mathbf{H}^{1}})}}{\sigma_{\mathbf{H}^{1}}}$ and $C$ is a constant term which is ignored during optimization.

\section{Datasets}
\label{App: Datasets}

In this section we provide details on the datasets used during training.
The detailed statistics are summarized in Table \ref{tab: data stats}

\noindent \textbf{Molecular Interaction Prediction.}
For the datasets used in the molecular interaction prediction task, we convert the SMILES string into graph structure by using the Github code of CIGIN \cite{CIGIN}.
Moreover, for the datasets that are related to solvation free energies, i.e., MNSol, FreeSolv, CompSol, Abraham, and CombiSolv, we use the SMILES-based datasets provided in the previous work \cite{CombiSolv}. Only solvation free energies at temperatures of 298 K ($\pm$ 2) are considered and ionic liquids and ionic solutes are removed \cite{CombiSolv}.
\begin{itemize}[leftmargin=5mm]
    \item \textbf{Chromophore} \cite{Chromophore} contains 20,236 combinations of 7,016 chromophores and 365 solvents which are given in the SMILES string format. All optical properties are based on scientific publications and unreliable experimental results are excluded after examination of absorption and emission spectra.
    In this dataset, we measure our model performance on predicting \textbf{maximum absorption wavelength (Absorption)}, \textbf{maximum emission wavelength (Emission)} and \textbf{excited state lifetime (Lifetime)} properties which are important parameters for the design of chromophores for specific applications.
    We delete the NaN values to create each dataset which is not reported in the original scientific publications.
    Moreover, for Lifetime data, we use log normalized target value since the target value of the dataset is highly skewed inducing training instability.
    \item \textbf{MNSol} \cite{MNSol} contains 3,037 experimental free energies of solvation or transfer energies of 790 unique solutes and 92 solvents. In this work, we consider 2,275 combinations of 372 unique solutes and 86 solvents following previous work \cite{CombiSolv}.
    \item \textbf{FreeSolv} \cite{FreeSolv} provides 643 experimental and calculated hydration free energy of small molecules in water. In this work, we consider 560 experimental results following previous work \cite{CombiSolv}.
    \item \textbf{CompSol} \cite{CompSol} dataset is proposed to show how solvation energies are influenced by hydrogen-bonding association effects. We consider 3,548 combinations of 442 unique solutes and 259 solvents in the dataset following previous work \cite{CombiSolv}.
    \item \textbf{Abraham} \cite{Abraham} dataset is a collection of data published by the Abraham research group at College London. We consider 6,091 combinations of 1,038 unique solutes and 122 solvents following previous work \cite{CombiSolv}.
    \item \textbf{CombiSolv} \cite{CombiSolv} contains all the data of MNSol, FreeSolv, CompSol, and Abraham, resulting in 10,145 combinations of 1,368 solutes and 291 solvents.
\end{itemize}

\noindent \textbf{Drug-Drug Interaction Prediction.}
For the datasets used in the drug-drug interaction prediction task, we use the positive drug pairs given in MIRACLE Github link\footnote{\url{https://github.com/isjakewong/MIRACLE/tree/main/MIRACLE/datachem}}, which removed the data instances that cannot be converted into graphs from SMILES strings.
Then, we generate negative counterparts by sampling a complement set of positive drug pairs as the negative set for both datasets.
We also follow the graph converting process of MIRACLE \cite{MIRACLE} for classification task.
\begin{itemize}[leftmargin=5mm]
    \item \textbf{ZhangDDI} \cite{ZhangDDI} contains 548 drugs and 48,548 pairwise interaction data and multiple types of similarity information about these drug pairs.
    \item \textbf{ChChMiner} \cite{ChChMiner} contains 1,322 drugs and 48,514 labeled DDIs, obtained through drug labels and scientific publications.
\end{itemize}
Although ChChMiner dataset has much more drug instances than ZhangDDI dataset, the number of labeled DDI is almost the same. This indicates that ChChMiner dataset has much more sparse relationship between the drugs.

\begin{table}[t]
\small
    \centering
    \caption{Data statistics.}
    \begin{tabular}{c|cc|ccccc}
    \multicolumn{1}{c|}{Task} & \multicolumn{2}{c|}{Dataset}& $\mathcal{G}^{1}$ & $\mathcal{G}^{2}$ & \# $\mathcal{G}^{1}$ & \# $\mathcal{G}^{2}$ & \# Pairs \\ \hline \hline
    \multicolumn{1}{c|}{} & \multicolumn{1}{c|}{} & Absorption & Chrom. & Solvent & 6416 & 725 & 17276\\
    \multicolumn{1}{c|}{} & \multicolumn{1}{c|}{Chromophore \tablefootnote{\label{url: Chromophore} \url{https://figshare.com/articles/dataset/DB_for_chromophore/12045567/2}}} & Emission & Chrom. & Solvent & 6412 & 1021 & 18141\\
    \multicolumn{1}{c|}{} & \multicolumn{1}{c|}{} & Lifetime & Chrom. & Solvent & 2755 & 247 & 6960 \\ \cline{2-8}
    \multicolumn{1}{c|}{Molecular} & \multicolumn{2}{c|}{MNSol \tablefootnote{\url{https://conservancy.umn.edu/bitstream/handle/11299/213300/MNSolDatabase_v2012.zip?sequence=12&isAllowed=y}}} & Solute & Solvent & 372 & 86 & 2275 \\
    \multicolumn{1}{c|}{Interaction} & \multicolumn{2}{c|}{FreeSolv \tablefootnote{\url{https://escholarship.org/uc/item/6sd403pz}}} & Solute & Solvent & 560 & 1 & 560 \\
    \multicolumn{1}{c|}{} & \multicolumn{2}{c|}{CompSol \tablefootnote{\url{https://aip.scitation.org/doi/suppl/10.1063/1.5000910}}} & Solute & Solvent & 442 & 259 & 3548 \\
    \multicolumn{1}{c|}{} & \multicolumn{2}{c|}{Abraham \tablefootnote{\url{https://www.sciencedirect.com/science/article/pii/S0378381210003675}}} & Solute & Solvent & 1038 & 122 & 6091 \\
    \multicolumn{1}{c|}{} & \multicolumn{2}{c|}{CombiSolv \tablefootnote{\url{https://ars.els-cdn.com/content/image/1-s2.0-S1385894721008925-mmc2.xlsx}}} & Solute & Solvent & 1495 & 326 & 10145 \\ \hline
    \multicolumn{1}{c|}{Drug-Drug} & \multicolumn{2}{c|}{ZhangDDI \tablefootnote{\url{https://github.com/zw9977129/drug-drug-interaction/tree/master/dataset}}} & Drug & Drug & 544 & 544 & 40255 \\
    \multicolumn{1}{c|}{Interaction} & \multicolumn{2}{c|}{ChChMiner \tablefootnote{\url{http://snap.stanford.edu/biodata/datasets/10001/10001-ChCh-Miner.html}}} & Drug & Drug & 949 & 949 & 21082 \\ \hline
    \multicolumn{1}{c|}{Graph} & \multicolumn{2}{c|}{AIDS \tablefootnote{\label{url: Similarity} \url{https://github.com/yunshengb/SimGNN}}} & Mole. & Mole. & 700 & 700 & 490K \\
    \multicolumn{1}{c|}{Similarity} & \multicolumn{2}{c|}{IMDB \textsuperscript{\ref{url: Similarity}}} & Ego-net. & Ego-net. & 1500 & 1500 & 2.25M \\
    \multicolumn{1}{c|}{Learning} & \multicolumn{2}{c|}{OpenSSL \tablefootnote{\url{https://github.com/runningoat/hgmn_dataset}}} & Flow & Flow & 4308 & 4308 & 18.5M \\ \hline
    \end{tabular}
    \label{tab: data stats}
\end{table}

\noindent \textbf{Graph Similarity Learning.}
For graph similarity learning task, we use three commonly used datasets, i.e., AIDS, IMDB \cite{SimGNN}, and OpenSSL \cite{xu2017neural}.
\begin{itemize}[leftmargin=5mm]
    \item \textbf{AIDS} \cite{SimGNN} contains 700 antivirus screen chemical compounds and the labels that are related to the similarity information of all pair combinations, i.e., 490K labels. The labels are Graph Edit Distance (GED) scores which are computed with $A^{*}$ algorithm.
    \item \textbf{IMDB} \cite{SimGNN} contains 1,500 ego-networks of movie actors/actresses, where there is an edge if the two people appear in the same movie. Labels are related to the similarity information of all pair combinations, i.e., 2.25M labels. The labels are Graph Edit Distance (GED) scores which are computed with $A^{*}$ algorithm.
    \item \textbf{OpenSSL} \cite{xu2017neural} dataset is generated from popular open-source software OpenSSL\footnote{\url{https://www.openssl.org/}}, whose graphs denote the binary function's control flow graph. Labels are related to whether two binary functions are compiled from the same source code or not, since the binary functions that are compiled from the same source code are semantically similar to each other.
    In this work, we only consider the graphs that contain more than 50 nodes, i.e., OpenSSL [50, 200] setting in previous work \cite{H2MN}.
\end{itemize}

\section{Additional Experiments}

\subsection{Graph Similarity Learning}
\label{App: Graph Similarity Learning}
Graph similarity learning aims to approximate the function that measures the similarity between two graph entities, which is a long standing problem in graph theory.
Due to the exponential time complexity of traditional methods, e.g., Graph Edit Distance (GED) \cite{bunke1997relation} and Maximum Common Subgraph (MCS) \cite{bunke1998graph}, developing algorithmic approaches for measuring the similarity is crucial.
Recently, numerous approaches based on GNNs have recently been proposed. Specifically, SimGNN \cite{SimGNN} models the node- and graph-level interactions with histogram features and neural tensor networks, respectively.
Moreover, GMN \cite{GMN} considers the relationship between a graph pair through the cross-graph attention mechanism, and GraphSim \cite{GraphSim} leverages convolutional neural networks to extract the relationship of the paired graphs given a similarity matrix.
Most recently, $\text{H}^{2}\text{MN}$ \cite{H2MN} learns the similarity by matching hyperedges \cite{zhou2006learning,feng2019hypergraph}, which are regarded as subgraphs of the input graph.
Specifically, $\text{H}^{2}\text{MN}$ constructs hyperedges, and selects a number of certain hyperedges considering the pagerank values for each graph in the pair. Then, it models the correlation between the subgraphs with the complex cross-graph attention coefficients.

We verify the generality of \proposed~by conducting experiments on the similarity learning task~\cite{SimGNN,HGMN,H2MN}.
We evaluate the model performance in terms of MSE, Spearman's Rank Correlation Coefficient (denoted as $\rho$), and precision@10 (p@10)~\cite{H2MN} for the regression tasks, and AUROC for the classification task.
We observe that~\proposed~consistently outperforms $\text{H}^{2}\text{MN}$ on AIDS and IMDB datasets, whereas it performs competitively on OpenSSL dataset. 
We attribute this to the inherent characteristics of the datasets. That is, AIDS and IMDB datasets are chemical compounds and social networks, respectively, both of which intuitively contain core substructures. On the other hand, OpenSSL dataset consists of control flow graphs of binary functions in which determining the core substructures is non-trivial.
Considering that most real-world graphs, e.g., social networks, contain core substructures, we argue that~\proposed~is practical in reality.

\begin{table}[h]
    \small
    \centering
    \caption{Performance on similarity learning task.}
    \begin{tabular}{c|ccc|ccc|c}
               & \multicolumn{3}{c|}{AIDS} & \multicolumn{3}{c|}{IMDB} & OpenSSL \\ \cline{2-8}
               & MSE & $\rho$ & p@10 & MSE & $\rho$ & p@10 & AUROC \\ \hline \hline
    SimGNN     & $1.376$ & $0.824$ & $0.400$ & $1.264$ & $0.878$ & $0.759$ & $94.25$ \\
    GMN        & $4.610$ & $0.672$ & $0.200$ & $4.422$ & $0.725$ & $0.604$ & $93.91$ \\
    GraphSim   & $1.919$ & $0.849$ & $0.446$ & $0.743$ & $0.926$ & $0.828$ & $93.66$ \\
    HGMN       & $1.169$ & $\mathbf{0.905}$ & $0.456$ & $0.335$ & $0.919$ & $0.837$ & $95.87$ \\ \hline
    $\text{H}^2\text{MN}_{\text{RW}}$ & $0.936$ & $0.878$ & $0.496$ & $0.296$ & $0.918$ & $0.872$ & $92.21$\\
    $\text{H}^2\text{MN}_{\text{NE}}$ & $0.924$ & $0.883$ & $0.511$ & $0.297$ & $0.889$ & $0.875$ & $\mathbf{98.25}$ \\ \hline
    \proposed & $0.768$ & $0.899$ & $\mathbf{0.581}$ & $0.305$ & $0.931$ & $0.873$& $94.81$ \\
    \proposedcont & $\mathbf{0.760}$ & $0.898$ & $0.574$ & $\mathbf{0.289}$ & $\mathbf{0.934}$ & $\mathbf{0.877}$ & $95.49$ \\ \hline
    \end{tabular}
    \label{tab:similarity learning}
\end{table}

\subsection{On the selection of $\mathcal{G}^{1}$ and $\mathcal{G}^{2}$}
\label{App: Selection G1 G2}
The decision on which input graph should be set as $\mathcal{G}^1$ depends on the task. 
For general tasks, where paired entities have a similar impact on the target value, we can make the objective symmetric. 
However, we empirically observe that the symmetric and asymmetric objectives perform competitively as shown in Table \ref{tab: symmetric general}.

\begin{table}[h]
    \small
    \centering
    \caption{Comparisons between the symmetric loss and the asymmetric loss in drug-drug interaction prediction (AUROC) and graph similarity learning (MSE).}
    \begin{tabular}{c|c|cc|cc|cc}
     \multirow{4}{*}{Model}& \multirow{4}{*}{Loss} & \multicolumn{4}{c|}{\multirow{2}{*}{Drug-Drug Interaction}} & \multicolumn{2}{c}{Graph Similarity}\\
      &  & \multicolumn{2}{c}{} &  &  &  \multicolumn{2}{c}{Learning}\\ \cline{3-8}
      &  & \multicolumn{2}{c|}{Transductive} & \multicolumn{2}{c|}{Inductive} & \multirow{2}{*}{AIDS} & \multirow{2}{*}{IMDB} \\ \cline{3-6}
      & & ZhangDDI & ChChMiner & ZhangDDI & ChChMiner &  &  \\ \hline \hline
     \multirow{2}{*}{\proposed} & Symmetric & \textbf{94.74} & 98.26 & 73.09 & 80.49 & 0.775 & 0.321 \\
     & Asymmetric & 94.27 & \textbf{98.80} & \textbf{74.59} & \textbf{81.14} & \textbf{0.768} & \textbf{0.305} \\\hline
     \multirow{2}{*}{\proposedcont} & Symmetric & \textbf{94.68} & 98.54 & 74.17 & \textbf{81.51} & 0.762 & \textbf{0.266} \\
     & Asymmetric & 93.78 & \textbf{98.84} & \textbf{75.08} & 80.68 & \textbf{0.760} & 0.289 \\ \hline
    \end{tabular}
    \label{tab: symmetric general}
\end{table}

However, when domain knowledge exists (e.g., a chromophore (solute) has a significant impact on target value in molecular interaction), we intentionally used chromophore (solute) as $\mathcal{G}^1$, and solvent as $\mathcal{G}^2$.
In Table \ref{tab: symmetric domain}, we indeed observe that the model performance deteriorates severely when the symmetric objective is adopted.

\begin{table}[h]
    \small
    \centering
    \caption{Comparison to symmetric loss function in molecular interaction prediction.}
    \begin{tabular}{c|c|ccc}
    \multirow{3}{*}{Model} & \multirow{3}{*}{Loss} & \multicolumn{3}{c}{Molecular Interaction}\\ \cline{3-5}
     &  & \multicolumn{3}{c}{Chromophore}\\ \cline{3-5}
     &  & Absorption   & Emission     & Lifetime \\ \hline \hline
    \multirow{2}{*}{\proposed} & Symmetric  & $21.11 $ & $25.49 $ & $0.798 $ \\
                                 & Asymmetric & $ \mathbf{17.84} $ & $ \mathbf{24.44} $ & $ \mathbf{0.796} $ \\ \hline
    \multirow{2}{*}{\proposedcont} & Symmetric  & $20.13 $ & $24.81 $ & $0.795 $ \\
                                 & Asymmetric & $ \mathbf{18.11} $ & $ \mathbf{23.90} $ & $ \mathbf{0.771} $ \\ \hline
    \end{tabular}
    \label{tab: symmetric domain}
\end{table}

Therefore, when applying~\proposed, we suggest to consider the graph that is crucial to the task as $\mathcal{G}^{1}$ if there exists clear domain knowledge, and try both the symmetric and asymmetric losses, otherwise.

\subsection{Additional Model Analysis}
\label{App: Additional Model Analysis}

\begin{figure}[h]
    \centering
    \includegraphics[width= 0.8\textwidth]{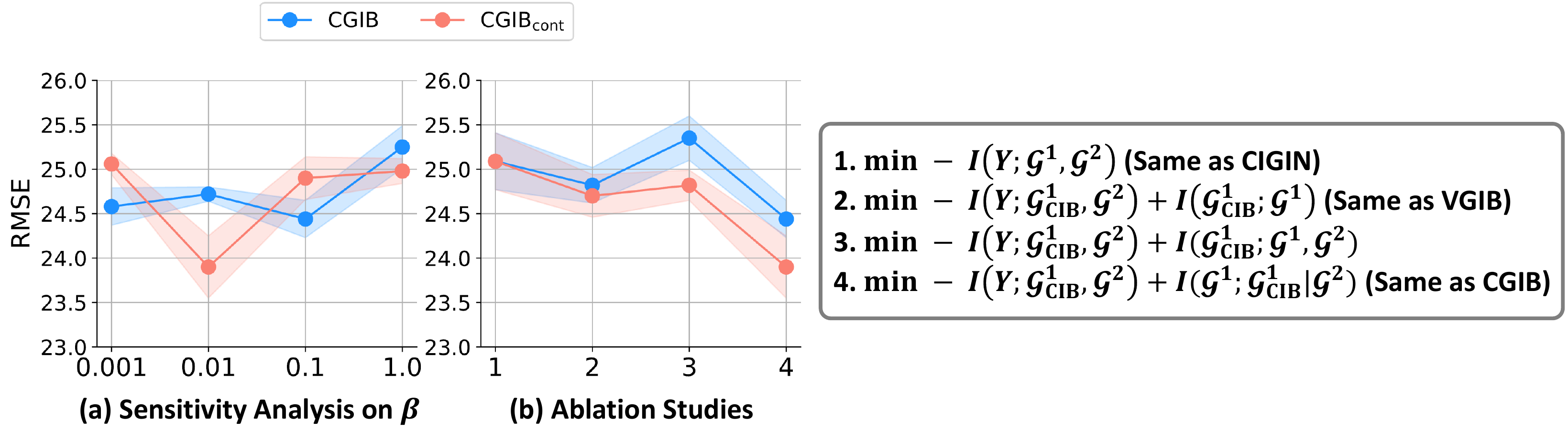}
    \caption{Model analysis on Emission dataset.}
    \label{fig: Model analysis on Emission}
\end{figure}

We provide sensitivity analysis and ablation studies results on Emission dataset in Figure \ref{fig: Model analysis on Emission}.

\subsection{Quantitative Analysis on CIB-Graph}
\label{App: Quantitative Analysis on CIB-Graph}

\begin{figure}[h]
    \centering
    \includegraphics[width= 0.4\textwidth]{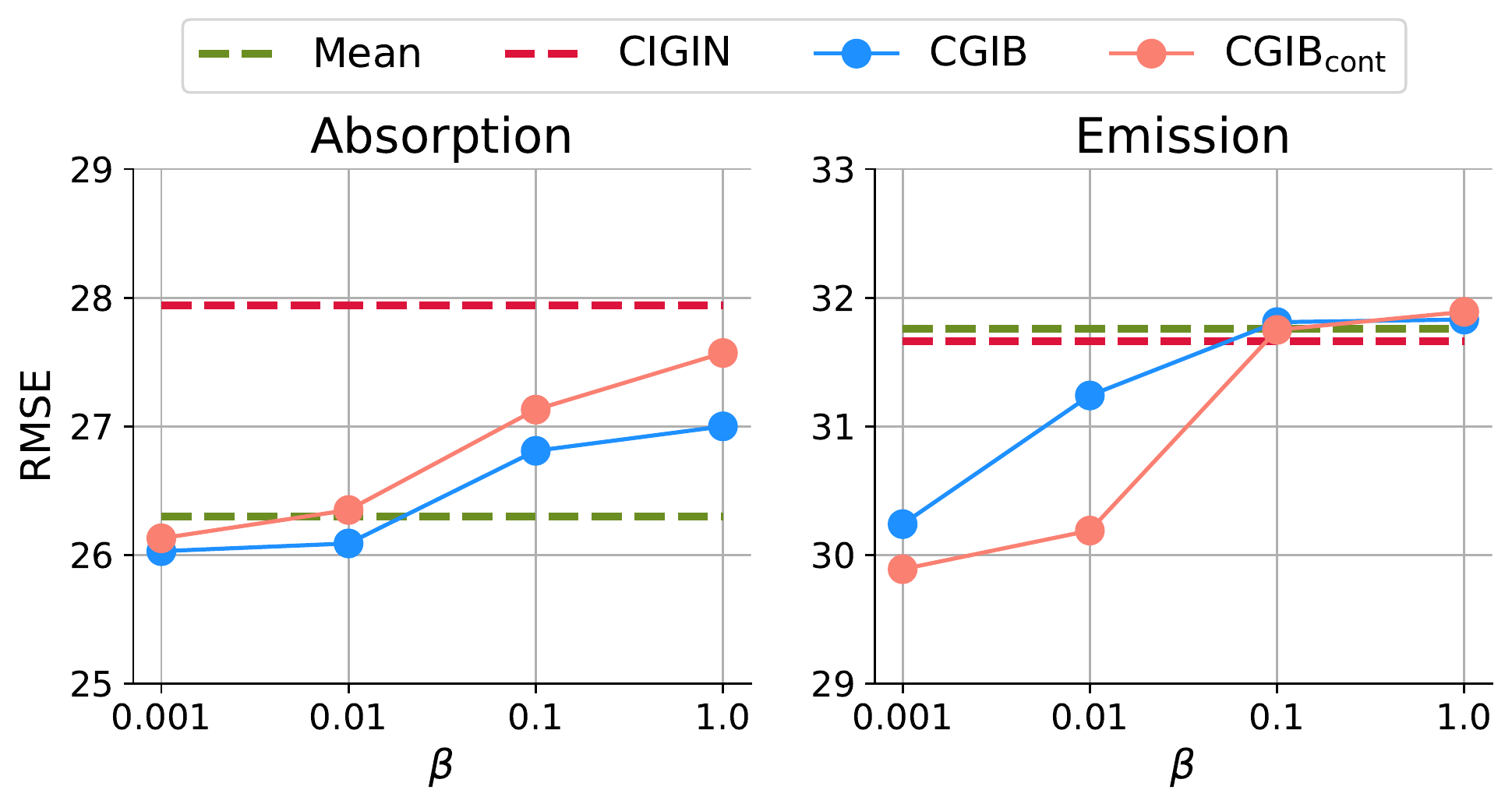}
    \caption{Quantitative analysis on CIB-Graph.}
    \label{fig: Quantiative Analysis CIB}
\end{figure}

{
To verify the usefulness of the subgraphs (i.e., $\mathcal{G}_{\mathrm{CIB}}^{1}$) determined by~\proposed~and~\proposedcont, we train a baseline model, i.e., MPNN, with the importance scores (i.e., $p_{i}$) obtained from~\proposed~and~\proposedcont.
Specifically, we replace the the graph readout function that is originally Set2Set~\cite{vinyals2015order} with the weighted sum pooling whose weights are based on the importance scores.
}
Similarly, we train another baseline model based on the attention scores obtained from the interaction map of CIGIN~\cite{CIGIN}.
Note that ``Mean'' indicates a baseline whose pooling function is Mean pooling, \textcolor{black}{i.e., utilizing all the nodes equally.} 
We have the following observations in Figure \ref{fig: Quantiative Analysis CIB}.
\textbf{1)} Attending to the subgraph determined by~\proposed~and~\proposedcont~outperforms the baseline model with Mean pooling function, indicating that our proposed models capture subgraphs that are useful for the task. Moreover, by comparing with the baseline model trained with subgraphs determined by CIGIN, we find out that~\proposed~and~\proposedcont~capture more useful subgraphs than CIGIN.
\textbf{2)} However, attending to the selected subgraphs does not always lead to performance improvements, i.e., when $\beta = 0.1$ and $\beta = 1.0$. 
{This is because aggressive compression (i.e., large $\beta$) impedes the detection of the task-relevant subgraph, which eventually degrades the model performance.}
This aligns with our observation in Section~\ref{sec: Sensitivity Analysis}.
\textbf{3)} Accordingly, the performance improves as $\beta$ decreases, because~\proposed~is trained to focus more on discovering the task-relevant part of the graph than aggressively compressing the graph.

\subsection{Sensitivity analysis on $I(\mathbf{Y} ; \mathcal{G}^{2})$}
\label{App: Sensitivity Analysis Gamma}

\begin{figure}[h]
    \centering
    \includegraphics[width=0.4\textwidth]{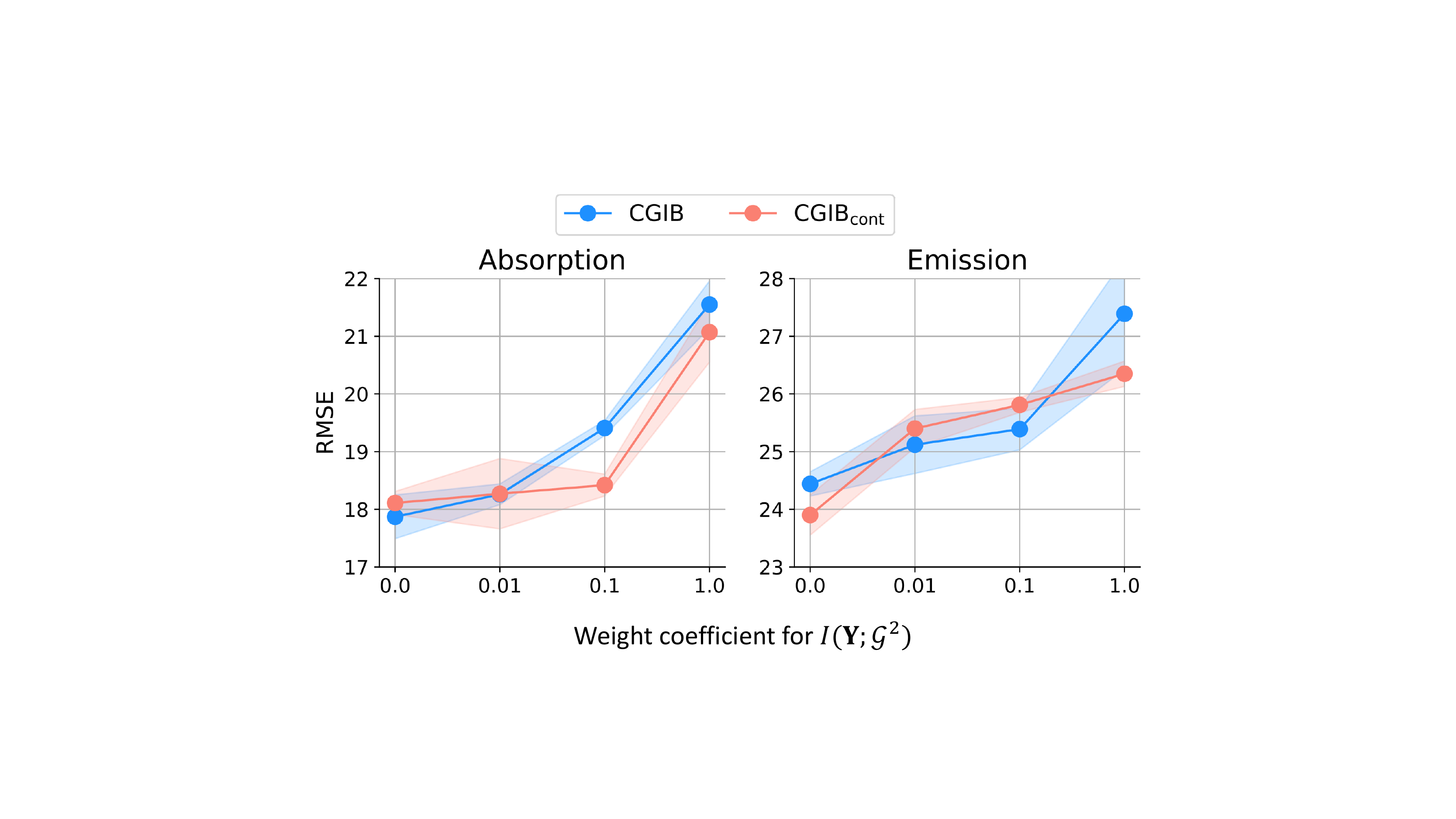}
    \caption{Sensitivity Analysis on $I(\mathbf{Y} ; \mathcal{G}^{2})$.}
    \label{fig: Sensitivity Gamma}
\end{figure}

Recall that the prediction term of \proposed~is decomposed as shown in Equation~\ref{eqn:decompose}:
$-I(\mathbf{Y} ;\mathcal{G}_{\mathrm{CIB}}^{1} | \mathcal{G}^{2}) = -I(\mathbf{Y} ;\mathcal{G}_{\mathrm{CIB}}^{1}, \mathcal{G}^{2}) + I(\mathbf{Y} ; \mathcal{G}^{2})$, and our goal is to minimize this term.
In this section, we analyze the effect of minimizing $I(\mathbf{Y} ; \mathcal{G}^{2})$ by conducting experiments on various weight coefficients for the term.
Specifically, we minimize the upper bound of $I(\mathbf{Y} ; \mathcal{G}^{2})$, which is given as the Kullback-Leibler divergence between $p(\mathbf{Y}|\mathcal{G}^{2})$ and its variational approximation $r(\mathbf{Y})$~\cite{VIB}.
Following VIB \cite{VIB}, we treat $r(\mathbf{Y})$ as a fixed spherical Gaussian, i.e., $r(\mathbf{Y}) = \mathcal{N}(\mathbf{Y}|0, 1)$.
As shown in Figure \ref{fig: Sensitivity Gamma}, we find out that including $I(\mathbf{Y} ; \mathcal{G}^{2})$ into our objectives severely deteriorates the model performance.
In other words, the model performs worse as the weight coefficient of $I(\mathbf{Y} ; \mathcal{G}^{2})$ gets larger, and the performance is the best when the weight coefficient is 0, i.e., when the term is not included in the optimization.
This is because minimizing the conditional mutual information $-I(\mathbf{Y} ;\mathcal{G}_{\mathrm{CIB}}^{1} | \mathcal{G}^{2})$ is equivalent to predicting the target value $\mathbf{Y}$ given $\mathcal{G}_{\mathrm{CIB}}^{1}$ conditioned on $\mathcal{G}^{2}$, i.e., the model focuses more on $\mathcal{G}_{\mathrm{CIB}}^{1}$ and do not fully use the information of $\mathcal{G}^{2}$.
Thus, we argue that in relational learning, where both entities in the pair are relevant to the target value $\mathbf{Y}$, considering only the joint mutual information $-I(\mathbf{Y} ;\mathcal{G}_{\mathrm{CIB}}^{1}, \mathcal{G}^{2})$ is more beneficial than additionally considering the conditional mutual information for prediction.

\subsection{Various modeling choices for $p_{\xi}$}
\label{App: Additional Experiments 1}
\begin{figure}[h]
    \centering
    \includegraphics[width= 0.4\textwidth]{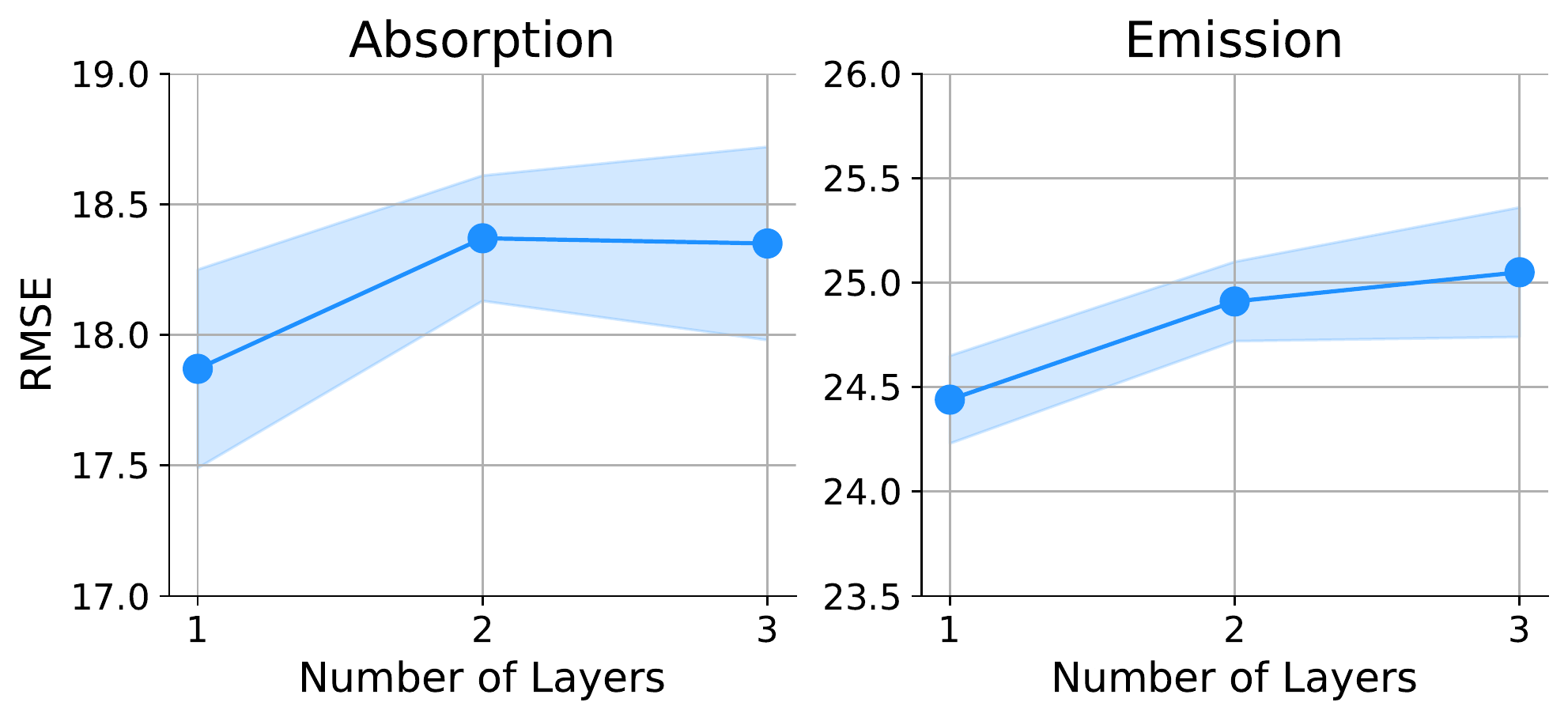}
    \captionof{figure}{Sensitivity analysis on the number of layers for $p_{\xi}$.}
    \label{fig: Sensitivity Solvent Layers}
\end{figure}

In Figure \ref{fig: Sensitivity Solvent Layers}, we show experimental results adopting more complex modeling choices for $p_{\xi}$ used in Equation~\ref{eq: bound3}. We find out that as the number of layers increases, the model performs worse, demonstrating that a complex modeling of $p_{\xi}$ incurs the information loss during the prediction of $\mathcal{G}^{2}$.

\subsection{Qualitative analysis for $\beta$}
\label{App: Additional Experiments 2}
\begin{figure}[h]
    \centering
    \includegraphics[width= 0.5\textwidth]{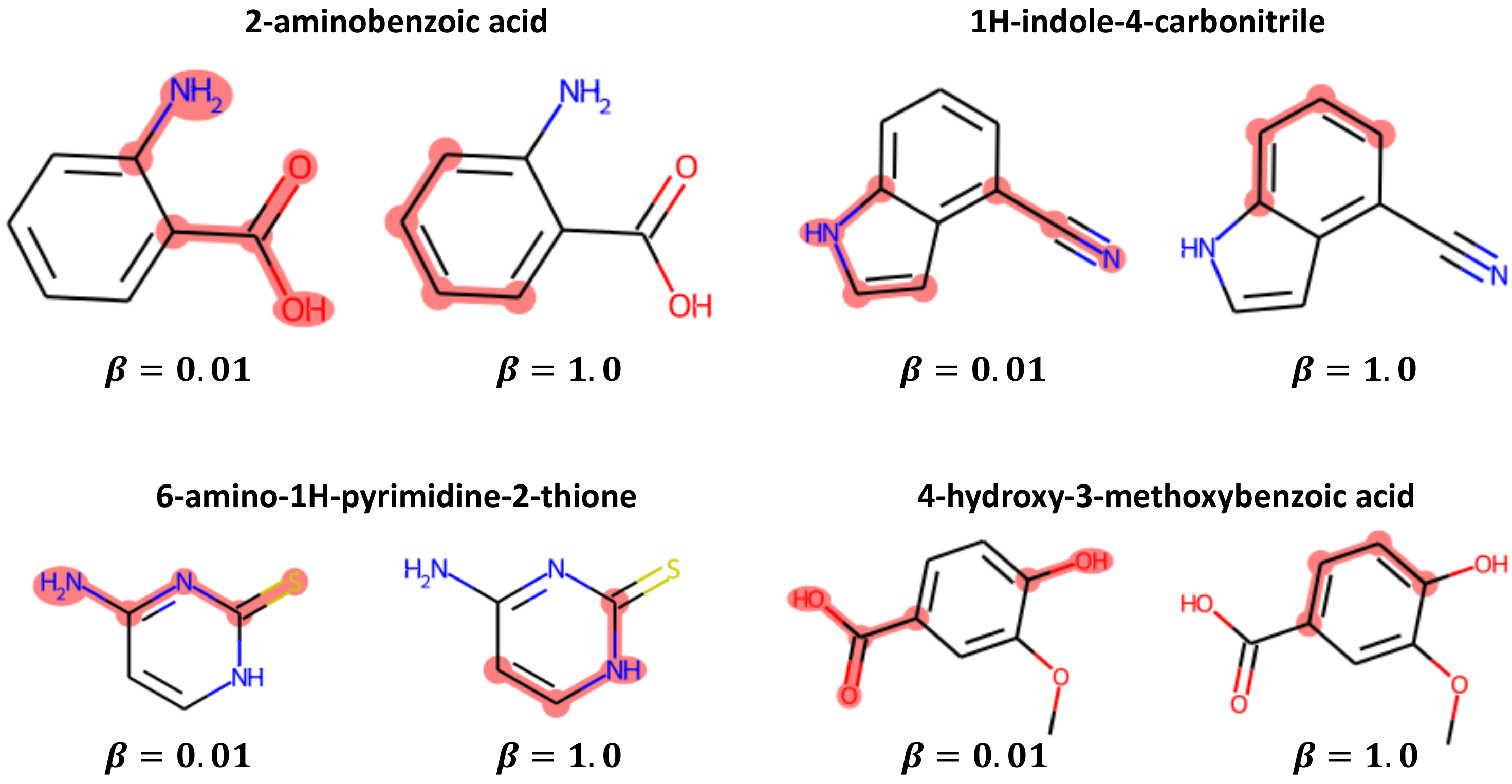}
    \captionof{figure}{Qualitative analysis for $\beta$.}
    \label{fig: Qualitative Beta}
\end{figure}

\looseness=-1
In Figure \ref{fig: Qualitative Beta}, we show how the model captures the core substructure according to $\beta$, which controls the trade-off between prediction and compression in our final objectives in Equation~\ref{eqn:finalloss}.
We observe that the model better captures task relevant substructures when $\beta = 0.01$ compared with the case when $\beta = 1.0$.
When $\beta = 1.0$, the model concentrates on the aromatic ring, which is the structure that makes the molecules stable rather than directly related to chemical reactions.
On the other hand, the model tends to discover the substructures related to chemical reactions when $\beta = 0.01$.
To sum up, as $\beta$ decreases, we find out that our model discovers more task-relevant but less compressed (i.e., large in scale) substructure of the molecule.

\section{Evaluation Protocol}
\label{App: Evaluation Protocol}
For the molecular interaction prediction task, we evaluate the models under 5-fold cross validation scheme following the previous work \cite{CIGIN}. The dataset is randomly split into 5 subsets and one of the subsets is used as the test set while the remaining subsets are used to train the model. A subset of the test set is selected as validation set for hyperparameter selection and early stopping.
We repeat 5-fold cross validation 5 times (i.e., 25 runs in total) and report the accuracy and standard deviation of the repeats.
For the DDI prediction task, we conduct experiments on both transductive and inductive settings. 
In the transductive setting, the graphs in test phase are also included in the training dataset.
That is, we use a random split of the data into train/validation/test data of 60/20/20\%\footnote{We make sure that all the graphs in validation and test data are seen during training.}, respectively, following SSI-DDI \cite{nyamabo2021ssi}.
On the other hand, in the inductive setting, the performance is evaluated when the models are presented with new graphs that were not included in the training dataset.
Specifically, let $\mathbb{G}$ denote the total set of graphs in the dataset. 
Given $\mathbb{G}$, we split $\mathbb{G}$ into $\mathbb{G}_{\mathrm{old}}$ and $\mathbb{G}_{\mathrm{new}}$, so that $\mathbb{G}_{\mathrm{old}}$ contains the set of graphs that have been seen in the training phase, and $\mathbb{G}_{\mathrm{new}}$ contains the set of graphs that have not been seen in the training phase.
Then, the new split of dataset consists of $\mathcal{D}_{\mathrm{train}} = \{ (\mathcal{G}^{1}, \mathcal{G}^{2}) \in \mathcal{D} | \mathcal{G}^{1} \in \mathbb{G}_{\mathrm{old}} \wedge \mathcal{G}^{2} \in \mathbb{G}_{\mathrm{old}} \}$ and $\mathcal{D}_{\mathrm{test}} = \{ (\mathcal{G}^{1}, \mathcal{G}^{2}) \in \mathcal{D} | (\mathcal{G}^{1} \in \mathbb{G}_{\mathrm{new}} \wedge \mathcal{G}^{2} \in \mathbb{G}_{\mathrm{new}}) \vee (\mathcal{G}^{1} \in \mathbb{G}_{\mathrm{new}} \wedge \mathcal{G}^{2} \in \mathbb{G}_{\mathrm{old}}) \vee (\mathcal{G}^{1} \in \mathbb{G}_{\mathrm{old}} \wedge \mathcal{G}^{2} \in \mathbb{G}_{\mathrm{new}}) \}$. 
We use a subset of $\mathcal{D}_{\mathrm{test}}$ as validation set in inductive setting.
For both the transductive and inductive DDI tasks, we repeat 5 independent experiments with different random seeds on a split data, and report the accuracy and the standard deviation of the repeats.
For the similarity learning task, we repeat 5 independent experiments with different random seeds on the already-split data given by \cite{H2MN}.
For all tasks, we report the test performance when the performance on the validation set gives the best result.

\section{Implementation Details}
\label{App: Implementation Details}
\textbf{Model architecture.} For the molecular interaction prediction, we use two 3-layer MPNNs \cite{MPNN} as our backbone molecule encoder to learn the representation of solute and solvent, respectively.
This is because the solute and solvent have different roles during the chemical reaction, thereby using a shared encoder may hard to capture the patterns in each solute and solvent.
On the other hand, we use a GIN \cite{GIN} to encode both drugs for the drug-drug interaction prediction task, and a GCN \cite{GCN} to encode both graphs for graph similarity learning.
Different from molecular interaction prediction, those tasks have no specific roles between two entities.
Therefore, we choose to share a encoder during the experiments.

\textbf{Model Training.} In all our experiments, we use the Adam optimizer for model optimization. 
For molecular interaction task and drug-drug interaction task, the learning rate was decreased on plateau by a factor of $10^{-1}$ with the patience of 20 epochs following previous work \cite{CIGIN}.
For similarity learning task, we do not use a learning rate scheduler for the fair comparison with $\text{H}^{2}\text{MN}$ \cite{H2MN}.

\noindent \textbf{Hyperparameter Tuning.} 
For fair comparisons, we follow the embedding dimensions and batch sizes of the state-of-the-art baseline for each task.
Detailed hyperparameter specifications are given in Table \ref{tab: Hyperparameters Specifications}.
For the hyperparameters of \proposed, we tune them in certain ranges as follows: learning rate $\eta$ in $\{5e^{-3}, 1e^{-3}, 5e^{-4}, 1e^{-4}, 5e^{-5}\}$ and $\beta$ in $\{1e^{-1}, 1e^{-2}, 1e^{-3}, 1e^{-4}, 1e^{-6}, 1e^{-8}, 1e^{-10} \}$. We also tune the temperature $\tau$ in $\{1.0, 0.5, 0.2\}$ for \proposedcont.
\begin{table}[h]
\small
\centering
\caption{Hyperparameter specifications ($*$: inductive task).}
    \begin{tabular}{c|c|c|c|cc|ccc}
     & Embedding & Batch & \multirow{2}{*}{Epochs} & \multicolumn{2}{c}{\proposed} & \multicolumn{3}{|c}{\proposedcont} \\ \cline{5-9}
     & Dim ($d$) & Size ($K$) & & lr & $\beta$ & lr & $\beta$ & $\tau$ \\ \hline \hline
     Absorption & 52 & 256 & 500 & 5e-3 & 1e-3 & 5e-3 & 1e-1 & 1.0\\
     Emission & 52  & 256 & 500 & 5e-3 & 1e-1 & 5e-3 & 1e-2 & 1.0\\
     Lifetime & 52  & 256 & 500 & 5e-3 & 1e-6 & 1e-3 & 1e-6 & 1.0\\
     MNSol & 42  & 32 & 200 & 1e-3 & 1e-4 & 1e-3 & 1e-6 & 1.0\\
     FreeSolv & 42  & 32 & 200 & 1e-3 & 1e-10 & 5e-3 & 1e-8 & 1.0\\
     CompSol & 42  & 256 & 500 & 1e-3 & 1e-8 & 1e-3 & 1e-6 & 1.0\\
     Abraham & 42  & 256 & 500 & 1e-3 & 1e-6 & 1e-3 & 1e-10 & 1.0\\
     CombiSolv & 42  & 256 & 500 & 1e-3 & 1e-4 & 5e-3 & 1e-6 & 0.5\\ \hline
     ZhangDDI & 300  & 512 & 500 & 5e-4 & 1e-3 & 5e-4 & 1e-3 & 0.5\\
     ChChMiner & 300  & 512 & 500 & 5e-4 & 1e-4 & 5e-4 & 1e-3 & 0.2\\ \hline
     $\text{ZhangDDI}^{*}$ & 300  & 512 & 500 & 5e-5 & 1e-4 & 5e-4 & 1e-4 & 1.0\\
     $\text{ChChMiner}^{*}$ & 300  & 512 & 500 & 5e-4 & 1e-4 & 5e-4 & 1e-4 & 1.0\\ \hline
     AIDS & 100 & 512 & 10000 & 1e-4 & 1e-3 & 1e-4 & 1e-4 & 1.0\\
     IMDB & 100 & 256 & 10000 & 1e-4 & 1e-3 & 1e-4 & 1e-3 & 1.0\\
     OpenSSL & 100 & 16 & 10000 & 1e-4 & 1e-6 & 1e-4 & 1e-4 & 1.0\\ \hline
    \end{tabular}
    \label{tab: Hyperparameters Specifications}
\end{table}

\textbf{Training Resources.}
We conduct all the experiments using a 24GB NVIDIA GeForce RTX 3090.




\end{document}